\documentclass[12pt,draftclsnofoot,onecolumn]{IEEEtran}

\usepackage{amsfonts}
\usepackage{setspace}
\usepackage{clrscode}
\usepackage{amsthm}
\usepackage{pict2e}

\usepackage{amsmath}
\usepackage{amssymb}
\usepackage[dvips]{graphicx}
\usepackage{epsfig}
\usepackage{hyperref}
\usepackage{algorithm}
\usepackage{algorithmic}
\usepackage{caption}
\usepackage{subcaption}
\usepackage{mathrsfs}

\newcommand{\rme}{{\mathrm{e}}}
\newcommand{\E}{\mathbb{E}}

\newcommand{\mH}{\mathcal{H}}
\newcommand{\hH}{\hat{\mathcal{H}}}

\newcommand{\figsize}{0.45}

\newcommand{\nid}{\rm{d}}
\newcommand{\f}{\rm{f}}

\newcommand{\mc}{\rm{c}}
\newcommand{\avg}{\text{avg}}
\newcommand{\pk}{\text{pk}}

\makeatletter

\newcommand{\Rmnum}[1]{\expandafter\@slowromancap\romannumeral #1@}
\makeatother

\newtheorem{Lem}{Theorem}
\newtheorem{Rem}{Remark}

\newtheorem{Prop}{Proposition}

\usepackage[ps2pdf,
bookmarks=false,
bookmarksnumbered=false, 
bookmarksopen=false, 
colorlinks=false]{}

\begin{document}

%
\title{Energy-Efficient Power Allocation in Cognitive Radio Systems with Imperfect Spectrum Sensing}

\author{
\IEEEauthorblockN{Gozde Ozcan, M. Cenk Gursoy, Nghi Tran, and Jian Tang}
\thanks{Gozde Ozcan, M. Cenk Gursoy and Jian Tang are with the Department of Electrical
Engineering and Computer Science, Syracuse University, Syracuse, NY, 13244
(e-mail: gozcan@syr.edu, mcgursoy@syr.edu, jtang02@syr.edu).}
\thanks{Nghi Tran is with the Department of Electrical and Computer Engineering, University of Akron, Akron, OH, 44325
(e-mail: nghi.tran@uakron.edu).}}

\maketitle

\vspace{-1cm}
\begin{abstract}
This paper studies energy-efficient power allocation schemes for secondary users in sensing-based spectrum sharing cognitive radio systems. It is assumed that secondary users first perform channel sensing possibly with errors and then initiate data transmission with different power levels based on sensing decisions. The circuit power is taken into account in total power consumption. In this setting, the optimization problem is to maximize energy efficiency (EE) subject to peak/average transmission power constraints and peak/average interference constraints. By exploiting quasiconcave property of EE maximization problem, the original problem is transformed into an equivalent parameterized concave problem and an iterative power allocation algorithm based on Dinkelbach's method is proposed. The optimal power levels are identified in the presence of different levels of channel side information (CSI) regarding the transmission and interference links at the secondary transmitter, namely perfect CSI of both transmission and interference links, perfect CSI of the transmission link and imperfect CSI of the interference link, imperfect CSI of both links or only statistical CSI of both links. Through numerical results, the impact of sensing performance, different types of CSI availability, and transmit and interference power constraints on the EE of the secondary users is analyzed.
\end{abstract}

\vspace{-.5cm}

\begin{spacing}{1.5}

\thispagestyle{empty}

\section{Introduction}
Driven by the emergence of new applications, growing demand for high data rates and increased number of users, energy consumption of wireless systems has gradually increased, leading to high levels of greenhouse gas emissions, consequent environmental concerns and high energy prices and operating costs. Therefore, optimal and efficient use of energy resources with the goal of reducing costs and minimizing the carbon footprint of wireless systems is of paramount importance, and energy-efficient design has become a vital consideration in wireless communications from the perspective of green operation. In addition, bandwidth bottleneck is a critical concern in wireless services since there are only limited spectral resources to accommodate multiple users and multiple wireless applications operating simultaneously. Despite this fact, the report released by the Federal Communication Commission showed that a large portion of the radio spectrum is not fully utilized most of the time \cite{fcc}. Hence, inefficiency in the spectrum usage opens the possibility for a new communication paradigm, called as cognitive radio \cite{mitola}, \cite{haykin}. In cognitive radio systems, unlicensed (cognitive or secondary) users are able to opportunistically access the frequency bands allocated to the licensed (primary) users as long as interference inflicted on primary users is limited below tolerable levels. Hence, cognitive radio leads to the dynamic and more efficient utilization of the available spectrum.

Motivated by the energy efficiency (EE) and spectral efficiency (SE) requirements and the need to address them jointly, we in this paper study EE in cognitive radio systems. More specifically, we investigate energy-efficient power allocation strategies in a practical setting with imperfect spectrum sensing.

\subsection{Literature Review}
EE in cognitive radio systems has been recently addressed in \cite{gur} - \cite{gabry}. For instance, the authors in \cite{gur} highlighted the benefits of cognitive radio systems for green wireless communications. The study in \cite{hong} mainly focused on SE and EE of cognitive cellular networks in 5G mobile communication systems. Also, the authors in \cite{pei} designed energy-efficient optimal sensing strategies and optimal sequential sensing order in multichannel cognitive radio networks. In addition, the sensing time and transmission duration were jointly optimized in \cite{shi}. Several recent studies investigated power allocation/control to maximize the EE in different settings. The authors in \cite{xiong} studied the optimal subcarrier assignment and power allocation to maximize either minimum EE among all secondary users or average EE in an OFDM-based cognitive radio network. Moreover, an optimal power loading algorithm was proposed in \cite{bedeer} to maximize the EE of an OFDM-based cognitive radio system in the presence of imperfect CSI of transmission link between secondary transmitter and secondary receiver. In the EE analysis of aforementioned works, secondary users are assumed to transmit only when the channel is sensed as idle. The work in \cite{wang} mainly focused on optimal power allocation to achieve the maximum EE of OFDM-based cognitive radio networks. Also, energy-efficient optimal power allocation in cognitive MIMO broadcast channel was studied in \cite{mao}. The authors in \cite{ramamonjison} proposed iterative algorithms to find the power allocation maximizing the sum EE of secondary users in heterogeneous cognitive two-tier networks. Additionally, the authors in \cite{gabry} studied the optimal power allocation and power splitting at the secondary transmitter that maximize the EE of the secondary user as long as a minimum secrecy rate for the primary user is satisfied. In these works, secondary users always share the spectrum with primary users without performing channel sensing.
\subsection{Main Contributions}
Unlike all the works mentioned in the previous subsection, in this study we consider a sensing-based spectrum sharing cognitive radio system in which secondary users can coexist with primary user under both idle and busy sensing decisions while adapting the transmission power according to the sensing results. In this setting, we determine the optimal power allocation schemes to maximize the EE of secondary users under constraints on both the transmission and interference power. The main contributions of this paper are summarized below:
\begin{itemize}

\item We formulate the EE maximization problem subject to peak/average transmit power constraints and peak/average interference constraint in the presence of imperfect sensing results. In particular, EE is defined as the ratio of the achievable rate over the total power expenditure including circuit power consumption. We explicitly show in the formulations the dependence of the achievable rate and energy efficiency on sensing performance and reliability.


\item After transforming the EE maximization problem into an equivalent concave form, we apply Dinkelbach's method to find the optimal power allocation schemes. Subsequently, we develop an efficient low-complexity power allocation algorithm for given channel fading coefficients of the transmission link between the secondary transmitter-receiver pair and of the interference link between the secondary transmitter and the primary receiver.

\item We study the impact of different levels of channel knowledge regarding the transmission and interference links on the optimal power allocation and the resulting EE performance of secondary users. In practice, perfect CSI is difficult to obtain due to the inherently time-varying nature of wireless channels, delay, noise and limited bandwidth in feedback channels and other factors. Therefore, our proposed power allocation schemes obtained under imperfect CSI are useful in practical settings and give insight about the impact of channel uncertainty on the system performance. At the same time, the EE in the presence of perfect CSI of the transmission and interference links serves as a baseline to compare the performances attained under the assumption of imperfect and statistical CSI of both links.


\item We also investigate the effects of imperfect sensing decisions, and the constraints on both transmit and interference power on the EE of secondary users.

\end{itemize}

The remainder of the paper is organized as follows. Section \ref{sec:system_model} introduces the system model and defines the EE of secondary users in the presence of imperfect sensing results. In Section \ref{sec:opt_power}, EE maximization problems subject to transmit and interference power constraints in the presence of imperfect sensing results and different levels of CSI regarding the transmission and interference links are formulated and the corresponding optimal power allocation schemes are derived. Subsequently, numerical results are presented and discussed in Section \ref{sec:num_results} before giving the main concluding remarks in Section \ref{sec:conc}.

\section{System Model} \label{sec:system_model}
We consider a sensing-based spectrum sharing cognitive radio system in which a secondary transmitter-receiver pair utilizes the spectrum holes in the licensed bands of the primary users. The term ``spectrum holes" denotes underutilized frequency intervals at a particular time and certain location. In order to detect the spectrum holes, secondary users initially perform channel sensing over a duration of $\tau$ symbols. It is assumed that secondary users employ frames of $T$ symbols. Hence, data transmission is performed in the remaining duration of $T-\tau$ symbols.

\subsection{Channel Sensing}
Spectrum sensing can be formulated as a hypothesis testing problem in which there are two hypotheses based on whether primary users are active or inactive over the channel, denoted by $\mH_{0}$ and $\mH_{1}$, respectively. Many spectrum sensing methods have been studied in the literature (see e.g., \cite{ghasemi}, \cite{axell} and references therein) including matched filter detection, energy detection and cyclostationary feature detection. Each method has its own advantages and disadvantages. However, all sensing methods are inevitably subject to errors in the form of false alarms and miss detections due to possibly low signal-to-noise ratio (SNR) levels of primary users, noise uncertainty, multipath fading and shadowing in wireless channels. Therefore, we consider that spectrum sensing is performed imperfectly with possible errors, and sensing performance depends on the sensing method only through detection and false alarm probabilities. As a result, any sensing method can be employed in the rest of the analysis. Let $\hH_1$  and $\hH_0$ denote the sensing decisions that the channel is occupied and not occupied by the primary users, respectively. Hence, by conditioning on the true hypotheses, the detection and false-alarm probabilities are defined, respectively, as follows:
\begin{align}
\mathscr{P}_{\nid} &= \Pr\{\hH_1 | \mH_1\}, \\
\mathscr{P}_{\f} &= \Pr\{\hH_1 | \mH_0\}.
\end{align}
Then, the conditional probabilities of idle sensing decision given the true hypotheses can be expressed as
\begin{align}
\Pr\{\hH_0 | \mH_1\}=1-\mathscr{P}_{\nid}, \\ Pr\{\hH_0 | \mH_0\}=1-\mathscr{P}_{\f}.
\end{align}
\subsection{Cognitive Radio Channel Model}
\begin{figure}[htb]
\centering
\includegraphics[width=0.5\textwidth]{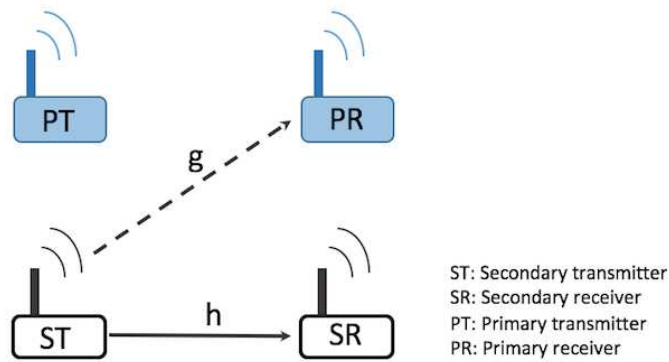}
\caption{System model.}
\label{fig:system_model}
\end{figure}

Following channel sensing, the secondary users initiate data transmission. The channel is considered to be block flat-fading channel in which the fading coefficients stay the same in one frame duration and vary independently from one frame to another. Secondary users are assumed to transmit under both idle and busy sensing decisions. Therefore, by considering the true nature of the primary user activity together with the channel sensing decisions, the four possible channel input-output relations between the secondary transmitter-receiver pair can be expressed as follows:
\begin{equation}
\begin{split}
y_i = \begin{cases} hx_{0,i}+ n_i & \text{if } (\mH_{0}, \hH_{0})\\
hx_{1,i}+ n_i  & \text{if } (\mH_{0}, \hH_{1}) \\
hx_{0,i}+ n_i + s_i  & \text{if } (\mH_{1}, \hH_{0}) \\
hx_{1,i}+ n_i + s_i & \text{if } (\mH_{1}, \hH_{1}),
\end{cases}
\end{split}
\label{eq:received_signa_BSl}
\normalsize
\end{equation}
where $i= 1, \dots, T-\tau$. Above, $x$  and $y$  are the transmitted and received signals, respectively and $h$ is the channel fading coefficient of the transmission link between the secondary transmitter and the secondary receiver, which is assumed to be Gaussian distributed with mean zero and variance $\sigma_h^2$. In addition, $n_i$ and $s_i$ denote the additive noise and the primary users' received faded signal. Both $\{n_i\}$ and $\{s_i\}$ are assumed to be independent and identically distributed circularly-symmetric, zero-mean Gaussian sequences with variances $N_0$ and $\sigma_s^2$, respectively. Moreover, the subscripts $0$ and $1$ in the transmitted signal indicate the transmission power levels of the secondary users. More specifically, the average power level is $P_0(g,h)$ if the channel is detected to be idle while it is $P_1(g,h)$ if the channel is detected to be busy. Also, $g$ denotes the channel fading coefficient of the interference link between the secondary transmitter and the primary receiver. System model is depicted in Fig. \ref{fig:system_model}.

Based on the input-output relation in (\ref{eq:received_signa_BSl}), the additive disturbance is given by
\begin{align} \label{eq:disturbance}
w_i=\begin{cases}  n_i &\text{ if } \mH_0 \text{ is true}  \\
n_i + s_i &\text{ if } \mH_1 \text{ is true}
\end{cases}.
\end{align}

In this setting, achievable rates of secondary users can be characterized by assuming Gaussian input and considering the input-output mutual information $I(x;y|h,\hH)$ given the fading coefficient $h$ and sensing decision $\hH$:
\begin{align} 
R_{G} &=\frac{T-\tau}{T}I(x;y|h,\hH) \label{eq:mutual_info_GM}
\\
\nonumber
&=\frac{T-\tau}{T}\bigg[\Pr\{\hH_0\}I(x_0;y|h,\hH_0)+\Pr\{\hH_1\}I(x_1;y|h,\hH_1)\bigg]
\\
\nonumber
&=\frac{T-\tau}{T} \Pr(\hH_0)\big[h(y|h,\hH_0)-h(y|x_0,h,\hH_0)\big] + \frac{T-\tau}{T} \Pr(\hH_1)(h(y|h,\hH_1)-h(y|x_1,h,\hH_1)), \\ \nonumber
&=\frac{T-\tau}{T} \Pr(\hH_0)\big[h(y|h,\hH_0)-h(w|h,\hH_0)\big] + \frac{T-\tau}{T} \Pr(\hH_1)\big[h(y|h,\hH_1)-h(w|h,\hH_1)\big],
\end{align}
where $h(.)$ denotes the differential entropy. Due to imperfect sensing results, the additive disturbance, $w$, follows a Gaussian mixture distribution and the differential entropy of Gaussian mixture density does not admit a closed-form expression. Hence, we do not have an explicit expression for $R_G$ with which we can identify the energy efficiency and obtain the optimal power allocation strategies. However, a closed-form achievable rate expression for secondary users can be obtained by replacing the Gaussian mixture noise with Gaussian noise with the same variance as shown in the next result.

\begin{Prop} \label{prop:1}
A closed-form achievable rate expression for secondary users in the presence of imperfect sensing decisions is given by
\begin{align}
\begin{split}\label{eq:lower_bound_R}
R_a&=\E_{g,h}\big\{R\big(P_0(g,h),P_1(g,h)\big)\big\} =\frac{T-\tau}{T}\sum_{k=0}^1\Pr(\hH_k)\E_{g,h}\Bigg\{\log\bigg(1+\frac{P_{k}(g,h)|h|^2}{N_0+\Pr(\mH_1|\hH_k)\sigma_s^2}\bigg)\Bigg\},
\end{split}
\normalsize
\end{align}
where $k \in \{0,1\}$ and $\E\{.\}$ denotes expectation operation. Also, $\Pr\{\hH_1\}$ and $\Pr\{\hH_0\}$ denote the probabilities of channel being detected as busy and idle, respectively, and can be expressed as
\begin{align}\label{eq:prob_idle_busy_sensed}
\Pr\{\hH_1\} &= \Pr\{\mH_0\}\mathscr{P}_{\f}+\Pr\{\mH_1\}\mathscr{P}_{\nid},\\
\Pr\{\hH_0\} &= \Pr\{\mH_0\}(1-\mathscr{P}_{\f})+\Pr\{\mH_1\}(1-\mathscr{P}_{\nid}).
\end{align}
\end{Prop}
\emph{Proof:} See Appendix \ref{appendix1}.

We note that since Gaussian is the worst-case noise \cite{medard}, the achievable rate expression $R_a$ in (\ref{eq:lower_bound_R}) is in general smaller than $R_G$ in (\ref{eq:mutual_info_GM}), which is the achievable rate obtained by considering the Gaussian-mixture noise. In the next result, we provide an upper bound on the difference between the two.

\begin{Lem} \label{teo:upper_bound}
The difference $(R_G-R_a)$ is upper bounded by
\begin{equation}
\begin{split} \label{eq:upper_bound1}
&R_G-R_a \le \Big(\frac{T-\tau}{T}\Big)\Bigg[ \E_{g,h}\Bigg\{\sum_{k=1}^2\log \Bigg(\frac{\sum_{i=1}^2\frac{\Pr (\mH_i|\hH_k)}{c_i}}{ \Big(1+\frac{P_k(g,h)|h|^2}{N_0 +  \Pr (H_1| \hH_k)\sigma_s^2}\Big)\sum_{i=1}^2\frac{\Pr (\mH_i | \hH_k)}{c_i + |h|^2P_k(g,h)}}\Bigg)\Bigg\} \\ &\hspace{2cm}-\sum_{k=1}^2 \Pr(\hH_k)\bigg(\frac{N_0 + \Pr (H_1 | \hH_k)\sigma_s^2}{N_0 + \sigma_s^2}\bigg) + \sum_{k=1}^2 \Pr(\hH_k) \E_{g,h}\bigg\{1+\frac{\Pr (\mH_1 | \hH_k)\sigma_s^2}{N_0 + |h|^2P_k(g,h)}\bigg\}\Bigg]
\end{split}
\normalsize
\end{equation}
where $c_1=N_0 + \sigma_s^2$ and $c_2 = N_0$.
\end{Lem}
\emph{Proof:} See Appendix \ref{appendix2}.

\begin{figure}[htb]
\centering
\includegraphics[width=0.5\textwidth]{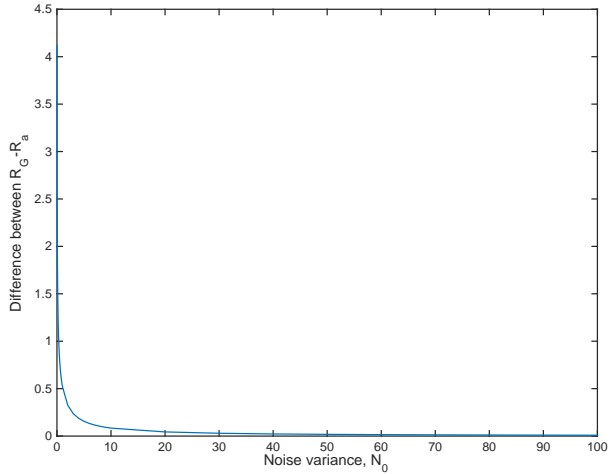}
\caption{Upper bound on the difference $(R_G-R_a)$ vs. noise variance $N_0$. }
\label{fig:difference}
\end{figure}

In Fig. \ref{fig:difference}, we plot the upper bound on the difference $(R_G-R_a)$ as a function of noise variance, $N_0$. We assume that $\mathscr{P}_{\nid}=0.8, \mathscr{P}_{\f}=0.1$ and $\sigma_s^2=1$. It is seen that as noise variance, $N_0$, increases, the upper bound approaches zero, hence $R_a$, which can be regarded as a lower bound on the achievable rate $R_G$, becomes tighter.

With the characterization of the achievable rate in (\ref{eq:lower_bound_R}), we now define the EE as the ratio of the achievable rate to the total power consumption:
\begin{align} \label{eq:achievable_EE}
\eta_{\rm{EE}}=\frac{\E_{g,h}\big\{R\big(P_0(g,h),P_1(g,h)\big)\big\}}{\E_{g,h}\{\Pr\{\hH_0\}P_0(g,h)+\Pr\{\hH_1\}P_1(g,h)\}+P_{\mc}}.
\end{align}
Above, the total power consists of average transmission power and circuit power, denoted by $P_{\mc}$. Circuit power represents the power consumed by the transmitter circuitry (i.e., by mixers, filters, and digital-to-analog converters, etc.), which is independent of transmission power.

The achievable EE expression in (\ref{eq:achievable_EE}) can serve as a lower bound since the lower bound on achievable rate $R_a$ in (\ref{eq:lower_bound_R}) is employed. The usefulness of this EE expression is due to its being an explicit function of the transmission power levels and sensing performance e.g., through the probabilities $\Pr\{\hH_0\}$, $\Pr\{\hH_1\}$, and $\Pr(\mH_1|\hH_k)$.
\begin{figure}[htb]
\centering
\includegraphics[width=0.5\textwidth]{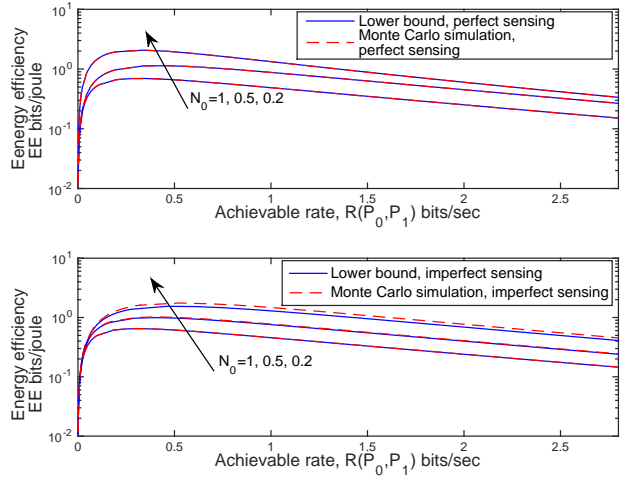}
\caption{Achievable EE $\eta_{\rm{EE}}$ vs. achievable rate $R_a$.}
\label{fig:EE_SE}
\end{figure}

In Fig. \ref{fig:EE_SE}, we plot the EE expression in (\ref{eq:achievable_EE}) (indicated as the lower bound) and the exact EE, in which we use Gaussian input and consider Gaussian mixture noise in the mutual information, as a function of achievable rate for both perfect sensing (i.e., $\mathscr{P}_{\nid} = 1$ and  $\mathscr{P}_{\f} = 0$) and imperfect sensing (i.e., $\mathscr{P}_{\nid} = 0.8$ and  $\mathscr{P}_{\f} = 0.2$). The graph is displayed in logarithmic scale to highlight the difference between the exact EE and the lower bound on EE. In order to evaluate the exact EE achieved with Gaussian input, we performed Monte Carlo simulations with $2 \times 10^6$ samples. In the case of perfect sensing, the lower bound and simulation result perfectly match as expected since in this case additive disturbance has Gaussian distribution rather than a Gaussian mixture. In the case of imperfect sensing, it is seen that the gap between the lower bound and exact EE decreases as $N_0$ increases, which matches with the characterization in Theorem \ref{teo:upper_bound}. Additionally, since circuit power is taken into consideration with value $P_c=0.1$, EE vs. achievable rate curve is bell-shaped and is also quasiconcave. It is observed that maximum EE is achieved at nearly the same achievable rate for both lower bound and exact EE expressions.

In the following section, we derive power allocation schemes that maximize the EE of the secondary users in the presence of sensing errors, different combinations of transmit power and average interference power constraints, and different levels of CSI regarding the transmission and interference links.

\section{Optimal Power Allocation}\label{sec:opt_power}
\subsection{Average Transmit Power Constraint and Average Interference Power Constraint}\label{subsec:average_constraints}

In this subsection, we obtain the optimal power allocation strategies to maximize the EE of secondary users under average transmit power and average interference power constraints in the presence of different levels of CSI regarding the transmission and interference links, namely perfect CSI of both transmission and interference links, perfect CSI of the transmission link and imperfect CSI of the interference link, imperfect CSI of both links, or statistical CSI of both links.

\subsubsection{Perfect CSI of both transmission and interference links} In this case, it is assumed that CSI of both transmission and interference links is perfectly known by the secondary transmitter. In this setting, the maximum EE under both average transmit power and interference power constraints can be found by solving the following optimization problem:
\begin{align}
\label{eq:EE_avg1}
&\max_{
\substack{P_0(g,h) \\ P_1(g,h)}} \eta_{\rm{EE}}\!=\! \frac{\E_{g, h}\big\{R\big(P_0(g,h),P_1(g,h)\big)\big\}}{\E_{g,h}\{\Pr\{\hH_0\}P_0(g,h)\!+\!\Pr\{\hH_1\}P_1(g,h)\}\!+\!P_{\mc}}\\ \label{eq:Pavg_cons1}
&\text{subject to} \hspace{0.4cm}\E_{g,h}\{\Pr\{\hH_0\}\,P_0(g,h)  + \Pr\{\hH_1\}\,P_1(g,h)\}  \le P_{\avg} \\ \label{eq:Qavg_cons1}
&\hspace{2.1cm}\E_{g,h}\{\big[(1-\mathscr{P}_{\nid})\,P_0(g,h) + \mathscr{P}_{\nid} \,P_1(g,h)\big]|g|^2\} \le Q_{\avg}\\
&\hspace{2.1cm}P_0(g,h)\geq 0 , P_1(g,h)\geq 0,
\end{align}
\normalsize
where $P_{\avg}$ denotes the maximum average transmission power of the secondary transmitter and $Q_{\avg}$ represents the maximum allowed average interference power at the primary receiver. In particular, average transmit power constraint in (\ref{eq:Pavg_cons1}) is chosen to satisfy the long-term power budget of the secondary users and average interference power constraint in (\ref{eq:Qavg_cons1}) is imposed to limit the interference, and hence to protect the primary user transmission. In this setting, the optimal power allocation strategy that maximizes the EE of secondary users is determined in the following result.

\begin{Lem} \label{teo:1} The optimal power allocation under the constraints in (\ref{eq:Pavg_cons1}) and (\ref{eq:Qavg_cons1}) is given by
\begin{align} \label{eq:opt_P0_avg1}
&P^*_0(g,\!h)=\Bigg[\!\frac{\frac{T-\tau}{T}\Pr\{\hH_0\}\log_2e}{(\lambda_1\!+\!\alpha)\Pr\{\hH_0\}+\nu_1 |g|^2(1-\mathscr{P}_{\nid})} -\frac{N_0\!+\!\Pr(\mH_1|\hH_0)\sigma_s^2}{|h|^2}\Bigg]^+\\\label{eq:opt_P1_avg1}
& P^*_1(g,\!h)=\Bigg[\frac{\frac{T-\tau}{T}\Pr\{\hH_1\}\log_2e}{(\lambda_1+\alpha)\Pr\{\hH_1\}+\nu_1 |g|^2\mathscr{P}_{\nid}}-\frac{N_0+\Pr(\mH_1|\hH_1)\sigma_s^2}{|h|^2}\Bigg]^+,
\end{align}
\normalsize
where $[x]^+$ denotes $\max(x,0)$, $\alpha$ is a nonnegative parameter, and $\lambda_1$ and $\nu_1$ are nonnegative Lagrange multipliers.
\end{Lem}

\emph{Proof:} See Appendix \ref{appendix3}.

Above, the Lagrange multipliers $\lambda_1$ and $\nu_1$ can be jointly obtained by inserting the optimal power allocation schemes (\ref{eq:opt_P0_avg1}) and (\ref{eq:opt_P1_avg1}) into the constraints (\ref{eq:Pavg_cons1}) and (\ref{eq:Qavg_cons1}). However, solving these constraints does not give closed-form expressions for $\lambda_1$ and $\nu_1$. Therefore, we employ the subgradient method, i.e., $\lambda_1$ and $\nu_1$ are updated iteratively according to the subgradient direction until convergence as follows:
\begin{align}
&\lambda_1^{(n+1)}=\Big[\lambda_1^{(n)}-t\big( P_{\avg}-\E_{g,h}\{\Pr\{\hH_0\}\,P_0^{(\!n\!)}(g,\!h) + \Pr\{\hH_1\}\,P_1^{(n)}(g,h)\}\big)\Big]^+ \\ \label{eq:nu_update}
&\nu_1^{(n+1)}\!=\!\!\Big[\nu_1^{(n)}-t\big( Q_{\avg}-\E_{g,h}\{\big[(1-\mathscr{P}_{\nid})\,P_0^{(n)}(g,h) +\mathscr{P}_{\nid} \,P_1^{(n)}(g,h)\big]|g|^2\}\big)\Big]^+,
\end{align}
\normalsize
where $n$ and $t$ denote the iteration index and the step size, respectively. When the step size is chosen to be constant, it was shown that the subgradient method is guaranteed to converge to the optimal value within a small range \cite{boyd}.

For a given value of $\alpha$, the optimal power levels in (\ref{eq:opt_P0_avg1}),  (\ref{eq:opt_P1_avg1}) can be found until $F(\alpha) \le \epsilon$ is satisfied. Dinkelbach's method converges to the optimal solution at a superlinear convergence rate. The detailed proof of convergence can be found in \cite{schaible}. In the case of $F(\alpha)=0$ in (\ref{eq:cond_opt}), the solution is optimal otherwise $\epsilon$-optimal solution is obtained. In the following table, Dinkelbach method-based iterative power allocation algorithm for EE maximization under imperfect sensing is summarized.

\begin{algorithm}
    \caption{Dinkelbach method-based power allocation that maximizes EE of cognitive radio systems under both average transmit power and interference constraints}
    \begin{algorithmic}[1]
      \STATE Initialization: $\mathscr{P}_{\nid}=\mathscr{P}_{\nid,init}$, $\mathscr{P}_{\f}=\mathscr{P}_{\f,init}$, $\epsilon > 0$, $\delta > 0$, $t > 0$, $\alpha^{(0)}=\alpha_{\text{init}}$,  $\lambda_1^{(0)}=\lambda_{1,\text{init}}$, $\nu_1^{(0)}=\nu_{1,\text{init}}$
      \STATE $n \leftarrow 0$
      \REPEAT
       \STATE calculate $P_0^*(g,h)$ and $P_1^*(g,h)$ using (\ref{eq:opt_P0_avg1}) and (\ref{eq:opt_P1_avg1}), respectively;
	\STATE update $\lambda_1$ and $\nu_1$ using subgradient method as follows:
        \STATE $k \leftarrow 0$
      \REPEAT
       \STATE $\lambda_1^{(k+1)}=\Big[\lambda_1^{(k)}-t\big( P_{\avg}-\E_{g,h}\{\Pr\{\hH_0\}\!\,P_0^{(\!k\!)}(g,h)  + \Pr\{\hH_1\}\,P_1^{(k)}(g,h)\}\big)\Big]^+$
	\STATE $\nu_1^{(k+1)}=\Big[\nu_1^{(k)}-t\big( Q_{\avg}-\E_{g,h}\{[(1-\mathscr{P}_{\nid})\,P_0^{(\!k\!)}(g,h)\! +\! \mathscr{P}_{\nid} \!\,P_1^{(k)}(g,h)]|g|^2\}\big)\Big]^+$
\STATE $k \leftarrow k+1$
       \UNTIL{$|\nu_1^{(k)}\big( Q_{\avg}-\E_{g,h}\{[(1-\mathscr{P}_{\nid})\,P_0^{(\!k\!)}(g,h) + \mathscr{P}_{\nid} \!\,P_1^{(\!k\!)}(g,h)]|g|^2\}\big)| \hspace{-0.1cm}\le \delta$ and $|\lambda_1^{(k)}\big( P_{\avg}-\E\{\Pr\{\hH_0\}\,P_0^{(\!k\!)}(g,\!h)  + \Pr\{\hH_1\}\,P_1^{(k)}(g,h)\}\big)| \le \delta$}
\STATE $\alpha^{(n+1)}=\frac{\E_{g,h}\big\{R\big(P_0^*(g,h),P_1^*(g,h)\big)\big\}}{\E_{g,h}\{\Pr\{\hH_0\}P_0^*(g,h)\!+\!\Pr\{\hH_1\}P_1^*(g,h)\}\!+\!P_{\mc}}$
\STATE $n \leftarrow n+1$
       \UNTIL{$|F(\alpha^{(n)})| \le \epsilon$}
    \end{algorithmic}
  \end{algorithm}

Note that in the case of $\alpha=0$, EE maximization problem is equivalent to SE maximization. Therefore, setting $\alpha=0$ in (\ref{eq:opt_P0_avg1}) and (\ref{eq:opt_P1_avg1}) provides the optimal power allocation strategies that maximize the average achievable rate of secondary users.

\begin{Rem}
The power allocation schemes in (\ref{eq:opt_P0_avg1}) and (\ref{eq:opt_P1_avg1}) have the structure of water-filling policy with respect to channel power gain $|h|^2$ between the secondary transmitter and secondary receiver, but average transmit and average interference power constraints are not necessarily satisfied with equality in constrast to to the case of throughput maximization. In addition, the water level in this policy depends on the interference channel power gain $|g|^2$ between the secondary transmitter and the primary receiver, i.e., less power is allocated when the interference link has a higher channel gain.
\end{Rem}

\begin{Rem}
The proposed power allocation schemes in (\ref{eq:opt_P0_avg1}) and (\ref{eq:opt_P1_avg1}) depend on the sensing performance through detection and false alarm probabilities, $\mathscr{P}_{\nid}$ and $\mathscr{P}_{\f}$, respectively. When both perfect sensing, i.e., $\mathscr{P}_{\nid}=1$ and $\mathscr{P}_{\f}=0$, and SE maximization are considered, i.e., $\alpha$ is set to $0$, the power allocation schemes become similar to that given in \cite{kang}. However, in our analysis the secondary users have two power allocation schemes depending on the presence or absence of active primary users.
\end{Rem}

\subsubsection{Perfect CSI of transmission link and imperfect CSI of interference link}
In practice, it may be difficult to obtain perfect CSI of the interference link due to the lack of cooperation between secondary and primary users. In this case, the channel fading coefficient of the interference link can be expressed as
\begin{align}
g=\hat{g}+\tilde{g},
\end{align}
where $\hat{g}$ denotes the estimate of the channel fading coefficient and $\tilde{g}$ represents the corresponding estimate error. It is assumed that $\hat{g}$ and $\tilde{g}$ follow independent, circularly symmetric complex Gaussian distributions with mean zero and variances $1-\sigma_g^2$ and $\sigma_g^2$, respectively, i.e., $\hat{g}\sim \mathcal{N}(0,1-\sigma_g^2)$ and $\tilde{g}\sim \mathcal{N}(0,\sigma_g^2)$. Under this assumption, the average interference constraint can be written as
\begin{align}
\begin{split}
\hspace{-0.25cm}Q_{\avg} &\geq \E_{g,\hat{g},h}\{\big[(1-\mathscr{P}_{\nid})P_0(\hat{g},h)+\mathscr{P}_{\nid}P_1(\hat{g},h)\big]|g|^2\} \\
&=\E_{\hat{g},h}\{\big[(1-\mathscr{P}_{\nid})P_0(\hat{g},h)+\mathscr{P}_{\nid}P_1(\hat{g},h)\big](|\hat{g}|^2+|\tilde{g}|^2)\}\\
&=\E_{\hat{g},h}\{\big[(1-\mathscr{P}_{\nid})P_0(\hat{g},h)+\mathscr{P}_{\nid}P_1(\hat{g},h)\big](|\hat{g}|^2+\sigma_g^2)\},
\end{split}
\end{align}
where the power levels $P_0$ and $P_1$ are now expressed as functions of the estimate $\hat{g}$. Now, the optimal power allocation problem under the assumptions of perfect instantaneous CSI of the transmission link and imperfect instantaneous CSI of the interfence link can be formulated as follows:
\begin{align}
\label{eq:EE_avg_imperfectCSI}
&\max_{
\substack{P_0(\hat{g},h) \\ P_1(\hat{g},h)}} \eta_{\rm{EE}}= \frac{\E_{\hat{g},h}\big\{R\big(P_0(\hat{g},h),P_1(\hat{g},h)\big)\big\}}{\E_{\hat{g},h}\{\Pr\{\hH_0\}P_0(\hat{g},h)\!+\!\Pr\{\hH_1\}P_1(\hat{g},h)\}\!+\!P_{\mc}}\\ \label{eq:Pavg_cons2}
&\text{subject to} \hspace{0.15cm}\E_{\hat{g},h}\{\Pr\{\hH_0\}\,P_0(\hat{g},h)  + \Pr\{\hH_1\}\,P_1(\hat{g},h)\}  \le P_{\avg} \\ \label{eq:Qavg_cons2}
&\hspace{1.8cm}\E_{\hat{g},h}\{\big[(1-\mathscr{P}_{\nid})\,P_0(\hat{g},h) + \mathscr{P}_{\nid} \,P_1(\hat{g},h)\big](|\hat{g}|^2+\sigma_g^2)\} \le Q_{\avg} \\
&\hspace{1.8cm}P_0(\hat{g},h)\geq 0 , P_1(\hat{g},h)\geq 0
\end{align}
\normalsize
In the following result, we determine the optimal power allocation strategy in closed-form for this case.

\begin{Lem} \label{teo:2} The optimal power allocation subject to the constraints in (\ref{eq:Pavg_cons2}) and (\ref{eq:Qavg_cons2}) is obtained as
\begin{align} \label{eq:opt_P0_avg2}
&\hspace{-0.45cm}P^*_0(\hat{g},h)=\Bigg[\frac{\frac{T-\tau}{T}\Pr\{\hH_0\}\log_2e}{(\lambda_2+\alpha)\Pr\{\!\hH_0\!\}+\nu_2 (1-\mathscr{P}_{\nid})(|\hat{g}|^2+\sigma_g^2)}-\frac{N_0+\Pr(\mH_1|\hH_0)\sigma_s^2}{|h|^2}\Bigg]^{+}\\\label{eq:opt_P1_avg2}
& \hspace{-0.45cm} P^*_1(\hat{g},h)=\Bigg[\frac{\frac{T-\tau}{T}\Pr\{\hH_1\}\log_2e}{(\lambda_2+\alpha)\Pr\{\hH_1\}+\nu_2 \mathscr{P}_{\nid}( |\hat{g}|^2+\sigma_g^2)}-\frac{N_0+\Pr(\mH_1|\hH_1)\sigma_s^2}{|h|^2}\Bigg]^{+},
\end{align}
\normalsize
where $\lambda_2$ and $\nu_2$ are nonnegative Lagrange multipliers associated with the average transmit power in (\ref{eq:Pavg_cons2}) and average interference power constraints in (\ref{eq:Qavg_cons2}), respectively.
\end{Lem}
\emph{Proof:} We mainly follow the same steps as in the proof of Theorem \ref{teo:1}, but with several modifications due to imperfect knowledge of the interference link. More specifically, the KKT conditions now become
\begin{align} \label{eq:P0_Lagrange}
&\frac{\frac{T-\tau}{T}\Pr\{\hH_0\}|h|^2\log_2e}{N_0+\Pr(\mH_1|\hH_0)\sigma_s^2+P_0^*(\hat{g},h)|h|^2}-(\lambda_2+\alpha) \Pr\{\hH_0\} -\nu_2 (|\hat{g}|^2+\sigma_g^2)(1-\mathscr{P}_{\nid})=0 \\ \label{eq:P1_Lagrange}
&\hspace{-0.2cm}\frac{\frac{T-\tau}{T}\Pr\{\hH_1\}|h|^2\log_2e}{N_0+\Pr(\mH_1|\hH_1)\sigma_s^2+P_1^*(g,\!h)|h|^2}-(\lambda_2+\alpha)  \Pr\{\hH_1\} -\nu_2 (|\hat{g}|^2+\sigma_g^2)\mathscr{P}_{\nid}=0 \\
&\lambda_2 (\E_{\hat{g},h}\{\Pr\{\hH_0\}\,P_0^*(\hat{g},h)  + \Pr\{\hH_1\}\,P_1^*(\hat{g},h)\}  - P_{\avg}) =0\\
& \nu_2 (\E_{\hat{g},h}\{\big[(1-\mathscr{P}_{\nid})\,P_0^*(\hat{g},h) + \mathscr{P}_{\nid} \,P_1^*(\hat{g},h)\big](|\hat{g}|^2+\sigma_g^2)\} -Q_{\avg})=0 \\
& \lambda_2 \geq 0, \nu_2 \geq 0.
\end{align}
\normalsize
Solving for $P^*_0(\hat{g},h)$ in (\ref{eq:P0_Lagrange}) and $P^*_1(\hat{g},h)$ in (\ref{eq:P1_Lagrange}) lead to the optimal power values in (\ref{eq:opt_P0_avg2}) and (\ref{eq:opt_P1_avg2}), respectively. \hfill $\square$

\begin{Rem}
We note that the optimal power levels in (\ref{eq:opt_P0_avg2}) and (\ref{eq:opt_P1_avg2}) now depend on the channel estimation error of the interference link, $\sigma_g^2$. More specifically, the water level is also determined by $\sigma_g^2$, i.e., inaccurate estimation with higher channel estimation error results in lower water levels, hence lower transmission powers.
\end{Rem}
\begin{Rem}
We can readily obtain the power allocation schemes under perfect sensing by setting  $\mathscr{P}_{\nid}=1$ and $\mathscr{P}_{\f}=0$ in (\ref{eq:opt_P0_avg2}) and (\ref{eq:opt_P1_avg2}). In addition, the proposed power schemes capture the power levels under perfect CSI of the interference link as a special case when $\sigma_g^2=0$.
\end{Rem}

\subsubsection{Imperfect CSI of both transmission and interference links}
In this case, we assume that in addition to the imperfect knowledge of the interference link, the secondary transmitter has imperfect CSI of the transmission link. The channel fading coefficient of the transmission link is written as
\begin{align}
h=\hat{h}+\tilde{h}.
\end{align}
Above, $\hat{h}$ is the estimate of the channel fading coefficient of the transmission link and $\tilde{h}$ is the corresponding estimation error. It is assumed that $\hat{h}$ and $\tilde{h}$ are independent, circularly symmetric complex Gaussian distributed with zero mean and variances $1-\sigma_h^2$ and $\sigma_h^2$, respectively, i.e., $\hat{h}\sim \mathcal{N}(0,1-\sigma_h^2)$ and $\tilde{h}\sim \mathcal{N}(0,\sigma_h^2)$. In this case, by taking into account imperfect CSI of both links, the achievable rate of secondary users is given by
\begin{align}
\begin{split}\label{eq:lower_bound_R_imperfect}
\hspace{-0.35cm}R_a&=\E_{\hat{g},\hat{h},h}\big\{R\big(P_0(\hat{g},\hat{h}),P_1(\hat{g},\hat{h})\big)\big\} \\ &=\!\frac{T-\tau}{T}\sum_{k=0}^1\Pr(\hH_k)\!\int_{\hat{g}}\!\bigg(\!\int_{\hat{h}}\!\bigg(\!\int_{|h|^2}\!\log\bigg(1+\frac{P_{k}(\eta,\zeta)\gamma }{N_0+\Pr(\mH_1|\hH_k)\sigma_s^2}\bigg) f_{|h|^2|\hat{h}}(\gamma \mid \hat{h})d\gamma \bigg)  f_{\hat{h}}(\zeta)d\zeta \bigg) f_{\hat{g}}(\eta)d\eta,
\end{split}
\normalsize
\end{align}
where $f_{|h|^2|\hat{h}}(\gamma \mid \hat{h})$ denotes the probability density function (pdf) of $|h|^2$ conditioned on $\hat{h}$, and in the case of Rayleigh fading, the corresponding pdf is given by
\begin{align}
f_{|h|^2|\hat{h}}(\gamma \mid \hat{h})=\frac{1}{\alpha_h^2}\rme^{-\frac{\gamma+|\hat{h}|^2}{\alpha_h^2}}I_0\bigg(\frac{2}{\alpha_h^2}\sqrt{|\hat{h}|^2\gamma}\bigg),
\end{align}
where $I_0(.)$ represents the modified Bessel function of the first kind \cite{abramowitz}. Consequently, the optimal power allocation problem can be expressed as
\begin{align}
\label{eq:EE_avg}
&\max_{
\substack{P_0(\hat{g},\hat{h}) \\ P_1(\hat{g},\hat{h})}} \eta_{\rm{EE}}\!=\! \frac{\E_{\hat{g},\hat{h},h}\big\{R\big(P_0(\hat{g},\hat{h}),P_1(\hat{g},\hat{h})\big)\big\}}{\E_{\hat{g},\hat{h},h}\{\Pr\{\hH_0\}P_0(\hat{g},\hat{h})\!+\!\Pr\{\hH_1\}P_1(\hat{g},\hat{h})\}\!+\!P_{\mc}} \\
\label{eq:Pavg_cons3}
&\text{subject to} \hspace{0.15cm}\E_{\hat{g},\hat{h},h}\{\Pr\{\hH_0\}\,P_0(\hat{g},\hat{h})  + \Pr\{\hH_1\}\,P_1(\hat{g},\hat{h})\}  \le P_{\avg} \\ \label{eq:Qavg_cons3}
& \hspace{1.8cm}\E_{\hat{g},\hat{h},h}\{\big[(1-\mathscr{P}_{\nid})\,P_0(\hat{g},\hat{h}) + \mathscr{P}_{\nid} \,P_1(\hat{g},\hat{h})\big](|\hat{g}|^2+\sigma_g^2)\} \le Q_{\avg}\\
&\hspace{1.8cm}P_0(\hat{g},\hat{h})\geq 0 , P_1(\hat{g},\hat{h})\geq 0
\end{align}
\normalsize
We obtain the following result for the optimal power allocation scheme.
\begin{Lem} \label{teo:3} The optimal power allocation subject to the constraints in (\ref{eq:Pavg_cons3}) and (\ref{eq:Qavg_cons3}) is obtained as
\begin{align} \label{eq:opt_power_P0_imperfect_both}
P^*_0(\hat{g},\hat{h})&=\bar{P}_0(\hat{g},\hat{h}) \\  \label{eq:opt_power_P1_imperfect_both}
P^*_1(\hat{g},\hat{h})&=\bar{P}_1(\hat{g},\hat{h}),
\end{align}
where $\bar{P}_0(\hat{g},\hat{h})$ and $\bar{P}_1(\hat{g},\hat{h})$ are the solutions, respectively, to the equations in (\ref{eq:P0_imperfectCSI}) and (\ref{eq:P1_imperfectCSI}) given on the next page. If there are no positive solutions for (\ref{eq:P0_imperfectCSI}) and (\ref{eq:P1_imperfectCSI}) given the values of $\hat{g}$ and $\hat{h}$, the instantaneous power levels are set to zero, i.e., $P^*_0(\hat{g},\hat{h})=0$ and $P^*_1(\hat{g},\hat{h})=0$.
\end{Lem}
\begin{figure*}
\begin{align} \label{eq:P0_imperfectCSI}
&\int_{0}^{\infty} \frac{\big(\frac{T-\tau}{T}\big)\Pr\{\hH_0\}\log_2e\gamma}{N_0+\Pr\{\hH_0\}\sigma_s^2+\bar{P}_0(\hat{g},\hat{h})\gamma}f_{|h|^2|\hat{h}}(\gamma,\hat{h})d\gamma=(\lambda_3+\alpha)\Pr\{\hH_0\}+\nu_3(1-\mathscr{P}_{\nid})(|\hat{g}|^2+\sigma_g^2) \\ \label{eq:P1_imperfectCSI}
&\int_{0}^{\infty} \frac{\big(\frac{T-\tau}{T}\big)\Pr\{\hH_1\}\log_2e\gamma}{N_0+\Pr\{\hH_1\}\sigma_s^2+\bar{P}_1(\hat{g},\hat{h})\gamma}f_{|h|^2|\hat{h}}(\gamma,\hat{h})d\gamma=(\lambda_3+\alpha)\Pr\{\hH_1\}+\nu_3 \mathscr{P}_{\nid}(|\hat{g}|^2+\sigma_g^2)
\end{align}
\hrule
\end{figure*}
\emph{Proof:} Similar to the proof of Theorem \ref{teo:1}, we first express the optimization problem in a subtractive form, which is a concave function of transmission power levels, and then define the Lagrangian as
\begin{equation} \label{eq:Lagrange_imperfect_both}
\small
\begin{split}
&L(P_0,P_1,\lambda_3,\nu_3,\alpha)=\E_{\hat{g},\hat{h},h}\big\{R\big(P_0(\hat{g},\hat{h}),P_1(\hat{g},\hat{h})\big)\big\}-\alpha (\E_{\hat{g},\hat{h},h}\{\Pr\{\hH_0\}P_0(\hat{g},\hat{h})+\Pr\{\hH_1\}P_1(\hat{g},\hat{h})\}\!+\!P_{\mc})\\&-\lambda_3 (\E_{\hat{g},\hat{h},h}\{\Pr\{\hH_0\}\,P_0(\hat{g},\hat{h})  + \Pr\{\hH_1\}\,P_1(\hat{g},\hat{h})\} -P_{\avg} )-\nu_3 (\E_{\hat{g},\hat{h},h}\{\big[(1-\mathscr{P}_{\nid})\,P_0(\hat{g},\hat{h}) \!+\! \mathscr{P}_{\nid} \,P_1(\hat{g},\hat{h})\big](|\hat{g}|^2\!+\!\sigma_g^2)\} \!-\!Q_{\avg}).
\end{split}
\end{equation}
\normalsize
Setting the derivatives of the Lagrangian in (\ref{eq:Lagrange_imperfect_both}) with respect to $P^*_0(\hat{g},\hat{h})$ and $P^*_1(\hat{g},\hat{h})$ to zero and arranging the terms yield the desired results in (\ref{eq:P0_imperfectCSI}) and (\ref{eq:P1_imperfectCSI}), respectively.
 \hfill $\square$

\begin{Rem}
Let $f_0(P_0(\hat{g},\hat{h}))$ and $f_1(P_1(\hat{g},\hat{h}))$ denote the left-hand sides of (\ref{eq:P0_imperfectCSI}) and (\ref{eq:P1_imperfectCSI}), respectively, as a function of the transmission powers and let $\omega_0$ and $\omega_1$ denote the right-hand sides of (\ref{eq:P0_imperfectCSI}) and (\ref{eq:P1_imperfectCSI}), respectively. For given values of $\hat{g}$ and $\hat{h}$, $f_0(P_0(\hat{g},\hat{h}))$ and $f_1(P_1(\hat{g},\hat{h}))$ are positive decreasing functions of transmission powers with their maximum values $f_0(0)$ and $f_1(0)$ obtained at $P_0(\hat{g},\hat{h})=0$ and $P_1(\hat{g},\hat{h})=0$, respectively. Hence, the optimal solutions $P^*_0$ and $P^*_1$ can be characterized as
\begin{align}
P^*_0=\begin{cases} f^{-1}_0(\omega_0) & 0<\omega_0<f_0(0) \\
0 &\omega_0 \geq f_0(0)
\end{cases} \\
P^*_1=\begin{cases} f^{-1}_1(\omega_1) & 0<\omega_1<f_1(0) \\
0 &\omega_1 \geq f_1(0).
\end{cases}
\end{align}
It is seen from the above expressions that we allocate power only when $f_0(0)>\omega_0$ and $f_1(0)>\omega_1$, otherwise power levels are zero. Also, the average transmission powers attained with the proposed above optimal power levels are decreasing functions of $\omega_1$ and $\omega_2$, respectively since $f^{-1}_0(\omega_0)$ and $f^{-1}_1(\omega_1)$ are decreasing in $\omega_0$ and $\omega_1$, respectively. Hence, in that sense, the optimal power allocation can be interpreted again as water-filling policy.
\end{Rem}

\begin{Rem}
The proposed power levels in (\ref{eq:opt_power_P0_imperfect_both}) and (\ref{eq:opt_power_P1_imperfect_both}) are functions of the variance of the estimation error of the transmission link, $\sigma_h^2$ and interference link, $\sigma_g^2$. Hence, Theorem \ref{teo:3} can be seen as a generalization of the power allocation schemes attained under perfect CSI of transmission link and interference links, i.e., this case can be recovered by setting $\sigma_h^2=0$ and $\sigma_g^2=0$.
\end{Rem}

\subsubsection{Statistical CSI of both transmission and interference links}

In this case, the secondary transmitter has only statistical CSI of both transmission and interference links, i.e., knows only the fading distribution of both transmission and interference links. Under this assumption, the power allocation problem is formulated as follows:
\begin{align}
\label{eq:EE_avg_statistical}
&\max_{
\substack{P_0, P_1 }} \eta_{\rm{EE}}\!=\! \frac{R(P_0,P_1)}{\E\{\Pr\{\hH_0\}P_0\!+\!\Pr\{\hH_1\}P_1\}\!+\!P_{\mc}}\\ \label{eq:Pavg_cons1_statistical}
&\text{subject to} \hspace{0.4cm}\E\{\Pr\{\hH_0\}\,P_0 + \Pr\{\hH_1\}\,P_1\}  \le P_{\avg} \\ \label{eq:Qavg_cons1_statistical}
&\hspace{1.6cm}\E\{\big[(1-\mathscr{P}_{\nid})\,P_0 + \mathscr{P}_{\nid} \,P_1\big]|g|^2\} \le Q_{\avg}\\
&\hspace{1.6cm}P_0\geq 0 , P_1\geq 0
\end{align}
\normalsize
Note that transmission power levels $P_0$ and $P_1$ are no longer functions of $g$ and $h$. There are no closed-form expressions for the optimal power levels $P^*_0$ and $P^*_1$. However, we can solve (\ref{eq:EE_avg_statistical}) numerically by transforming the optimization problem into an equivalent parametrized concave form and using convex optimization tools.

\subsection{Peak Transmit Power Constraint and Average Interference Power Constraint}
In this section, we assume that peak transmit power constraints are imposed rather than average power constraints. Interference is still controlled via average interference constraints. Peak transmit power constraint is imposed to limit the instantaneous transmit power of the secondary users, and hence corresponds to a stricter constraint compared to the average transmit power constraint.

\subsubsection{Perfect CSI of both transmission and interference links}
Energy-efficient power allocation under the assumption of perfect CSI of both transmission and interference links can be obtained by solving the following problem:
\begin{align}
\label{eq:EE_avg}
&\max_{
\substack{P_0(g,h) \\ P_1(g,h)}} \eta_{\rm{EE}}\!=\! \frac{\E_{g,h}\big\{R\big(P_0(g,h),P_1(g,h)\big)\big\}}{\E_{g,h}\{\Pr\{\hH_0\}P_0(g,h)\!+\!\Pr\{\hH_1\}P_1(g,h)\}\!+\!P_{\mc}}\\ \label{eq:P0_peak_perfect}
&\text{subject to} \hspace{0.35cm} P_0(g,h)  \le P_{\pk,0} \\  \label{eq:P1_peak_perfect}
&\hspace{2cm} P_1(g,h)  \le P_{\pk,1}
\end{align}
\begin{align} \label{eq:Qavg_peak_perfect}
&\E_{g,h}\{\big[(1-\mathscr{P}_{\nid})\,P_0(g,h) + \mathscr{P}_{\nid} \,P_1(g,h)\big]|g|^2\} \le Q_{\avg}\\
&P_0(g,h)\geq 0 , P_1(g,h)\geq 0,
\end{align}
\normalsize
where $P_{\pk,0}$ and $P_{\pk,1}$ denote the peak transmit power limits when the channel is detected as idle and busy, respectively. Under the above constraints, the optimal power allocation strategy is determined in the following result.
\begin{Lem} \label{teo:4} The optimal power allocation scheme that maximizes the EE of the secondary users subject to the constraints in (\ref{eq:P0_peak_perfect}), (\ref{eq:P1_peak_perfect}) and (\ref{eq:Qavg_peak_perfect}) is given by
\begin{equation} \label{P0_peak1}
\begin{split}
P^*_0(g,\!h) \!=\! \!\begin{cases} 0, &\hspace{-0.1cm} |g|^2 \geq \check{g}_{1,0}\\
\!\!\frac{\frac{T-\tau}{T}\Pr\{\hH_0\}\log_2e}{\nu_4 |g|^2(1-\mathscr{P}_{\nid})+\alpha\Pr\{\hH_0\}}\!\!-\!\!\frac{N_0+\Pr(\mH_1|\hH_0)\sigma_s^2}{|h|^2},  &\hspace{-0.1cm} \check{g}_{1,0}\!>\! |g|^2 \!\!>\! \check{g}_{2,0} \\
P_{\pk,0}, &\hspace{-0.1cm} |g|^2 \le \check{g}_{2,0}
\end{cases} \\
\end{split}
\normalsize
\end{equation}
\begin{equation}\label{P1_peak1}
\begin{split}
P^*_1(g,\!h) \!= \!\!\begin{cases} 0, &|g|^2 \geq \check{g}_{1,1}\\
\!\!\frac{\frac{T-\tau}{T}\Pr\{\hH_1\}\log_2e}{\nu_4 |g|^2 \mathscr{P}_{\nid}+\alpha \Pr\{\hH_1\}}\!-\!\frac{N_0+\Pr(\mH_1|\hH_1)\sigma_s^2}{|h|^2},  & \check{g}_{1,1}\!>\! |g|^2 \!\!>\! \check{g}_{2,1} \\
P_{\pk,1}, &|g|^2 \le  \check{g}_{2,1}
\end{cases}
\end{split}
\normalsize
\end{equation}
where
\begin{align}
\check{g}_{1,j}&=\frac{1}{\nu_4 \rho_j}\bigg( \frac{\frac{T-\tau}{T}\Pr\{\hH_j\}\log_2e |h|^2}{N_0+\Pr(\mH_1|\hH_j)\sigma_s^2}-\alpha\Pr\{\hH_j\}\bigg), \\
\check{g}_{2,j}&=\frac{1}{\nu_4 \rho_j} \bigg(\frac{\frac{T-\tau}{T}\Pr\{\hH_j\}\log_2e |h|^2}{P_{\pk,j}|h|^2+N_0+\Pr(\mH_1|\hH_j)\sigma_s^2}-\alpha\Pr\{\hH_j\}\bigg).
\end{align}
\normalsize
In the above expressions, $j \in \{0,1\}$, $\rho_0=1-\mathscr{P}_{\nid}$ and $\rho_1=\mathscr{P}_{\nid}$.
\end{Lem}
\emph{Proof:} By transforming the above optimization problem into an equivalent parametrized concave form and following the same steps as in the proof of Theorem \ref{teo:1} with peak transmit power constraints instead of average transmit power constraint, we can readily obtain the optimal power allocation schemes as in (\ref{P0_peak1}) and (\ref{P1_peak1}), respectively.

\begin{Rem}
Different from Theorem \ref{teo:1}, the optimal power levels are limited by $P_{pk,0}$ and $P_{pk,1}$, respectively, when the channel fading coefficient of the interference link is less than a certain threshold, which is mainly determined by the sensing performance through the detection and false-alarm probabilities.
\end{Rem}

\begin{Rem}
By setting $\alpha=0$, $\mathscr{P}_{\nid}=1$ and $\mathscr{P}_{\f}=0$ in (\ref{P0_peak1}) and (\ref{P1_peak1}), we can see that the power allocation schemes in (\ref{P0_peak1}) and (\ref{P1_peak1}) have similar structures as those in \cite{kang} in the case of throughput maximization where average interference power constraint is satisfied with equality. However, this constraint is not necessarily satistifed with equality in EE maximization.
\end{Rem}

Algorithm 1 can be modified to maximize the EE subject to peak power constraints and average interference constraint in such a way that $P_0^*(g,h)$ and $P_1^*(g,h)$ are computed using (\ref{P0_peak1}) and (\ref{P1_peak1}), respectively and only Lagrange multiplier $\nu_4$ is updated according to (\ref{eq:nu_update}).

\subsubsection{Perfect CSI of the transmission link and imperfect CSI of the interference link}
In the presence of perfect CSI of the transmission link and imperfect CSI of the interference link, the optimization problem in (\ref{eq:EE_avg_imperfectCSI}) is subject to
\begin{align} \label{eq:P0_peak_imperfect}
& P_0(\hat{g},h)  \le P_{\pk,0} \\  \label{eq:P1_peak_imperfect}
&P_1(\hat{g},h)  \le P_{\pk,1} \\  \label{eq:Qavg_peak_imperfect}
&\E_{\hat{g},h}\{\big[(1-\mathscr{P}_{\nid})\,P_0(\hat{g},h) + \mathscr{P}_{\nid} \,P_1(\hat{g},h)\big](|\hat{g}|^2+\sigma_g^2)\} \le Q_{\avg}
\end{align}
The main characterization for the optimal power allocation is as follows:

\begin{Lem} \label{teo:5} The optimal power allocation scheme under the constraints in (\ref{eq:P0_peak_imperfect}), (\ref{eq:P1_peak_imperfect}) and (\ref{eq:Qavg_peak_imperfect}) is obtained as
\begin{equation} \label{P0_peak_imperfect}
\small
\begin{split}
\hspace{-0.2cm}P^*_0(\hat{g},\!h) \!\!=\! \!\begin{cases} 0, &\hspace{-0.1cm} |\hat{g}|^2 \geq \hat{g}_{1,0}\\
\!\!\frac{\frac{T-\tau}{T}\Pr\{\hH_0\}\log_2e}{\nu_5 (|\hat{g}|^2\!+\sigma_g^2)(1\!-\!\mathscr{P}_{\nid})+\alpha\!\Pr\{\!\hH_0\!\}}\!\!-\!\!\frac{N_0\!+\Pr(\mH_1|\hH_0)\sigma_s^2}{|h|^2},  &\hspace{-0.1cm}\hat{g}_{1,0}\!>\!\! |\hat{g}|^2 \!\!>\! \hat{g}_{2,0} \\
P_{\pk,0}, &\hspace{-0.1cm}|\hat{g}|^2 \le \hat{g}_{2,0}
\end{cases} \\
\end{split}
\normalsize
\end{equation}
\begin{equation}\label{P1_peak_imperfect}
\small
\begin{split}
\hspace{-0.2cm}P^*_1(\hat{g},\!h) \!\!=\! \!\begin{cases} 0, &|\hat{g}|^2 \geq \hat{g}_{1,1}\\
\!\!\frac{\frac{T-\tau}{T}\Pr\{\hH_1\}\log_2e}{\nu_5 (|\hat{g}|^2\!+\sigma_g^2)\!\mathscr{P}_{\nid}+\alpha\!\Pr\{\hH_1\!\}}\!-\!\!\frac{N_0\!+\Pr(\mH_1|\hH_1)\sigma_s^2}{|h|^2},  &\hat{g}_{1,1}\!>\! |\hat{g}|^2 \!\!>\! \hat{g}_{2,1} \\
P_{\pk,1}, &|\hat{g}|^2 \le \hat{g}_{2,1}.
\end{cases}
\end{split}
\normalsize
\end{equation}
Above,
\small
\begin{align}
\hat{g}_{1,j}&=\frac{1}{\nu_5 \rho_j}\bigg( \frac{\frac{T-\tau}{T}\Pr\{\hH_j\}\log_2e |h|^2}{N_0+\Pr(\mH_1|\hH_j)\sigma_s^2}-\alpha\Pr\{\hH_j\}\bigg)-\sigma_g^2, \\
\hat{g}_{2,j}&=\frac{1}{\nu_5 \rho_j} \bigg(\frac{\frac{T-\tau}{T}\Pr\{\hH_j\}\log_2e |h|^2}{P_{\pk,j}|h|^2+N_0+\Pr(\mH_1|\hH_j)\sigma_s^2}-\alpha\Pr\{\hH_j\}\bigg)-\sigma_g^2.
\end{align}
\normalsize
\end{Lem}
Since similar steps as in the proof of Theorem \ref{teo:1} are followed, the proof is omitted.
\begin{Rem}
In the optimal power allocation schemes given in (\ref{P0_peak_imperfect}) and (\ref{P1_peak_imperfect}), the cut-off values, $\hat{g}_{1,j}$ and $\hat{g}_{2,j}$ for the estimated channel power gain of the interference link depend on the channel estimation error variance of the interference link as different from the results in Theorem $\ref{teo:4}$.
\end{Rem}

\subsubsection{Imperfect CSI of both transmission and interference links}
We again have peak transmit power and average interference power constraints. However, different from the previous subsections, the transmission link is imperfectly known at the secondary transmitter. Therefore, the power levels are functions of $\hat{g}$ and $\hat{h}$. We derive the following result for the optimal power allocation schemes that maximize the EE of the secondary users. Again the proof is omitted for brevity.
\begin{Lem} \label{teo:6} The optimal power allocation under peak transmit power and average interference power constraints is given by
\begin{align}
P^*_0(\hat{g},\hat{h})&=\min\big(P_{\pk,0},\bar{P}_0(\hat{g},\hat{h})\big)\\
P^*_1(\hat{g},\hat{h})&=\min\big(P_{\pk,1},\bar{P}_1(\hat{g},\hat{h})\big) ,
\end{align}
where $\bar{P}_0(\hat{g},\hat{h})$ is solution to
\begin{align}
\begin{split}
\hspace{-0.2cm}\int_{0}^{\infty}\frac{\big(\frac{T-\tau}{T}\big)\Pr\{\hH_0\}\log_2e\gamma}{N_0+\Pr\{\hH_0\}\sigma_s^2+\bar{P}_0(\hat{g},\hat{h})\gamma}f_{|h|^2|\hat{h}}(\gamma,\hat{h})d\gamma=\alpha\Pr\{\hH_0\} + \nu_6(1-\mathscr{P}_{\nid})(|\hat{g}|^2+\sigma_g^2)
\end{split}
\normalsize
\end{align}
and $\bar{P}_1(\hat{g},\hat{h})$ is solution to
\begin{align}
\begin{split}
\hspace{-0.2cm}\int_{0}^{\infty} \frac{\big(\frac{T-\tau}{T}\big)\Pr\{\hH_1\}\log_2e\gamma}{N_0+\Pr\{\hH_1\}\sigma_s^2+\bar{P}_1(\hat{g},\hat{h})\gamma}f_{|h|^2|\hat{h}}(\gamma,\hat{h})d\gamma=\alpha\Pr\{\hH_1\}+\nu_6 \mathscr{P}_{\nid}(|\hat{g}|^2+\sigma_g^2).
\end{split}
\normalsize
\end{align}
\end{Lem}

\subsubsection{Statistical CSI of both transmission and interference links}
In this case, the optimal power allocation problem is subject to peak transmit and average interference power constraints under the assumption of the availability of only statistical CSI of both transmission and interference links. The optimal values of $P^*_0$ and $P^*_1$ can be found numerically by converting the optimization problem into an equivalent parametrized concave form and employing convex optimization tools.

\subsection{Average Transmit Power Constraint and Peak Interference Power Constraint}
Finally, we consider the case in which the secondary transmitter operates under average transmit power constraint and peak interference power constraints, which are imposed to satisfy short-term QoS requirements of the primary users.

\subsubsection{Perfect CSI of both transmission and interference links}
In this case, the objective function in (\ref{eq:EE_avg1}) is subject to the following constraints:
\begin{align} \label{eq:Pavg_Qpk}
&\E_{g,h}\{\Pr\{\hH_0\}\,P_0(g,h)  + \Pr\{\hH_1\}\,P_1(g,h)\}  \le P_{\avg} \\  \label{eq:Pavg_Qpk1}
&P_0(g,h)|g|^2 \le Q_{\pk,0} \\  \label{eq:Pavg_Qpk2}
&P_1(g,h)|g|^2 \le Q_{\pk,1}
\end{align}
where $Q_{\pk,k}$ for $k \in \{0,1\}$ represents the peak limit on the received interference power at the primary receiver. Under these constraints, we derive the optimal power allocation scheme as follows:
\begin{Lem} \label{teo:7}
The optimal power allocation strategy under average transmit power in (\ref{eq:Pavg_Qpk}) and peak interference power constraints in (\ref{eq:Pavg_Qpk1}) and (\ref{eq:Pavg_Qpk2}) is obtained as
\begin{align} \label{eq:opt_P0_avg1_Qpk}
&P^*_0(g,\!h)=\min \Bigg( \Bigg[\frac{\frac{T-\tau}{T}\log_2e}{(\lambda_4+\alpha)} \!-\!\frac{N_0\!+\!\Pr(\mH_1|\hH_0)\sigma_s^2}{|h|^2}\Bigg]^+\!,\frac{Q_{pk,0}}{|g|^2}\Bigg)\\\label{eq:opt_P1_avg1_Qpk}
&P^*_1(g,\!h)=\min \Bigg( \Bigg[\frac{\frac{T-\tau}{T}\log_2e}{(\lambda_4+\alpha)} \!-\!\frac{N_0\!+\!\Pr(\mH_1|\hH_1)\sigma_s^2}{|h|^2}\Bigg]^+\!,\frac{Q_{pk,1}}{|g|^2}\Bigg)
\end{align}
Above, $\lambda_4$ is the Lagrange multiplier associated with the average transmit power in (\ref{eq:Pavg_Qpk}).
\end{Lem}
\emph{Proof:} The optimization problem is first expressed in terms of an equivalent concave form. Then, the similar steps as in the proof of Theorem \ref{teo:1} are followed. However, peak interference power constraints are imposed instead of average interference power constraint. Therefore, in this case the optimal powers are limited by peak interference power constraints, $Q_{\pk,k}$ for $k \in \{0,1\}$.

\subsubsection{Perfect CSI of the transmission link and imperfect CSI of the interference link}
We have the following constraints for the optimization problem in (\ref{eq:EE_avg_imperfectCSI}):
\begin{align}\label{eq:Pavg_Qpk_imperfect}
&\E_{g,h}\{\Pr\{\hH_0\}\,P_0(g,h)  + \Pr\{\hH_1\}\,P_1(g,h)\}  \le P_{\avg} \\  \label{eq:Pavg_Qpk1_imperfect}
&\Pr (P_0(g,h)|g|^2 \geq Q_{\pk,0}| \hat{g}) \le \xi_{0} \\  \label{eq:Pavg_Qpk2_imperfect}
&\Pr (P_1(g,h)|g|^2 \geq Q_{\pk,1}| \hat{g}) \le \xi_{1}
\end{align}
where $\xi_k$ for $k \in \{0,1\}$ denotes the outage threshold. Note that if the interference link CSI is imperfect, peak interference limit cannot be always satisfied and hence the peak interference constraints are modified as peak interference outage constraints. The constraints in (\ref{eq:Pavg_Qpk1_imperfect}) and (\ref{eq:Pavg_Qpk2_imperfect}) can be further expressed as \cite{sboui}
\begin{align}
P_0(g,h) \le \frac{Q_{\pk,0}}{F^{-1}_{|g|^2|\hat{g}}(1-\xi_0,\hat{g})} \\
P_1(g,h) \le \frac{Q_{\pk,1}}{F^{-1}_{|g|^2|\hat{g}}(1-\xi_1,\hat{g})}.
\end{align}
Above, $F^{-1}_{|g|^2|\hat{g}}(.,\hat{g})$ represents the inverse cumulative density function of $|g|^2$ given $\hat{g}$.
In this setting, the main characterization is given as follows:
\begin{Lem} \label{teo:8}
The optimal power allocation strategy under the constraints in (\ref{eq:Pavg_Qpk1_imperfect}) and (\ref{eq:Pavg_Qpk2_imperfect}) is given by
\begin{align} \label{eq:opt_P0_avg1_Qpk_imperfect}
&\hspace{-0.5cm}P^*_0(\hat{g},\!h)=\min \Bigg( \Bigg[\frac{\frac{T-\tau}{T}\log_2e}{(\lambda_5+\alpha)} \!-\!\frac{N_0\!+\!\Pr(\mH_1|\hH_0)\sigma_s^2}{|h|^2}\Bigg]^+\!\!\!,\frac{Q_{\pk,0}}{F^{-1}_{|g|^2|\hat{g}}(1-\xi_0,\hat{g})}\!\Bigg)\\\label{eq:opt_P1_avg1_Qpk_imperfect}
&\hspace{-0.5cm}P^*_1(\hat{g},\!h)=\min \Bigg( \Bigg[\frac{\frac{T-\tau}{T}\log_2e}{(\lambda_5+\alpha)} \!-\!\frac{N_0\!+\!\Pr(\mH_1|\hH_1)\sigma_s^2}{|h|^2}\Bigg]^+\!\!\!,\frac{Q_{\pk,1}}{F^{-1}_{|g|^2|\hat{g}}(1-\xi_1,\hat{g})}\!\Bigg)
\end{align}
\normalsize
Above, $\lambda_5$ is the Lagrange multiplier.
\end{Lem}
The proof is omitted for brevity.

\subsubsection{Imperfect CSI of both transmission and interference links}
It is assumed that the secondary users operate under the constraints below:
\begin{align}\label{eq:Pavg_Qpk_imperfect_both}
&\E_{g,h}\{\Pr\{\hH_0\}\,P_0(g,h)  + \Pr\{\hH_1\}\,P_1(g,h)\}  \le P_{\avg} \\  \label{eq:Pavg_Qpk1_imperfect_both}
&P_0(g,h) \le \frac{Q_{\pk,0}}{F^{-1}_{|g|^2|\hat{g}}(1-\xi_0,\hat{g})} \\  \label{eq:Pavg_Qpk2_imperfect_both}
&P_1(g,h) \le \frac{Q_{\pk,1}}{F^{-1}_{|g|^2|\hat{g}}(1-\xi_1,\hat{g})}.
\end{align}
In the following result, we derive the optimal power allocation scheme for this case.
\begin{Lem} \label{teo:7}
The optimal power allocation strategy under average transmit power in (\ref{eq:Pavg_Qpk}) and peak interference power constraints in (\ref{eq:Pavg_Qpk1}) and (\ref{eq:Pavg_Qpk2}) is obtained as
\begin{align}
P^*_0(\hat{g},\hat{h})&=\min\bigg(\frac{Q_{\pk,0}}{F^{-1}_{|g|^2|\hat{g}}(1-\xi_0,\hat{g})},\bar{P}_0(\hat{g},\hat{h})\bigg),\\
P^*_1(\hat{g},\hat{h})&=\min\bigg(\frac{Q_{\pk,1}}{F^{-1}_{|g|^2|\hat{g}}(1-\xi_1,\hat{g})},\bar{P}_1(\hat{g},\hat{h})\bigg)
\end{align}
where $\bar{P}_0(\hat{g},\hat{h})$ is solution to
\begin{align}
\begin{split}
\hspace{-0.2cm}\int_{0}^{\infty}\frac{\big(\frac{T-\tau}{T}\big)\Pr\{\hH_0\}\log_2e\gamma}{N_0+\Pr\{\hH_0\}\sigma_s^2+\bar{P}_0(\hat{g},\hat{h})\gamma}f_{|h|^2|\hat{h}}(\gamma,\hat{h})d\gamma=(\lambda_6 + \alpha)\Pr\{\hH_0\} \\
\end{split}
\normalsize
\end{align}
and $\bar{P}_1(\hat{g},\hat{h})$ is solution to
\begin{align}
\begin{split}
\hspace{-0.2cm}\int_{0}^{\infty} \frac{\big(\frac{T-\tau}{T}\big)\Pr\{\hH_1\}\log_2e\gamma}{N_0+\Pr\{\hH_1\}\sigma_s^2+\bar{P}_1(\hat{g},\hat{h})\gamma}f_{|h|^2|\hat{h}}(\gamma,\hat{h})d\gamma=(\lambda_6 + \alpha)\Pr\{\hH_1\}.
\end{split}
\normalsize
\end{align}
\end{Lem}
Again, the proof is omitted for the sake of brevity.

\section{Numerical Results}\label{sec:num_results}
In this section, we present numerical results to illustrate the EE of secondary users attained with the proposed EE maximizing power allocation methods in the presence of imperfect sensing results and different levels of CSI regarding the transmission and interference links. Unless mentioned explicitly, it is assumed that noise variance is $N_0=0.1$, the variance of primary user signal is $\sigma_s^2=1$. Also, the prior probabilities are $\Pr\{\mH_0\} = 0.4 $ and $\Pr\{\mH_1\} = 0.6$. The frame duration $T$ and sensing duration $\tau$ are set to $100$ and $10$, respectively. The circuit power is $P_c = 0.1$. The step sizes  $\lambda$ and $\nu$ are set to $0.1$ and tolerance $\epsilon$ is chosen as $10^{-6}$.

\begin{figure}[htb]
\centering
\includegraphics[width=\figsize\textwidth]{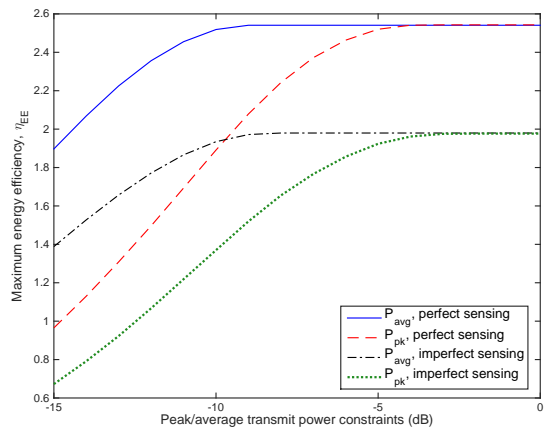}
\caption{Maximum EE $\eta_{\rm{EE}}$ vs. peak/average transmit power constraints.}
\label{fig:EE_transmitpower_limit}
\end{figure}

In Fig. \ref{fig:EE_transmitpower_limit}, we display maximum EE as a function of peak/average transmit power constraints for perfect sensing (i.e., $\mathscr{P}_{\nid}=1$ and $\mathscr{P}_{\f}=0$) and imperfect sensing (i.e., $\mathscr{P}_{\nid}=0.8$ and $\mathscr{P}_{\f}=0.1$). It is assumed that $Q_{\avg}=-8$ dB. Optimal power allocation is performed by assuming perfect instantaneous CSI at the secondary transmitter. It is seen that perfect detection of the primary user activity results in higher EE compared to the case with imperfect sensing decisions. In particular, the probabilities $\Pr(\mH_1|\hH_0)$ and $\Pr(\mH_0|\hH_1)$ are zero due to perfect spectrum sensing and hence, the secondary users do not experience additive disturbance from primary users, which leads to higher achievable rates, and hence higher EE compared to imperfect sensing case. It is also observed that maximum EE increases with increasing peak/average transmit power constraints. When peak/average transmit power constraints become sufficiently large compared to $Q_{\avg}$, maximum EE stays constant since the power is determined by average interference constraint, $Q_{\avg}$ rather than peak/average transmit power constraints. Moreover, higher EE is obtained under average transmit power constraint since the optimal power allocation under average transmit power constraint is more flexible than that under peak transmit power constraint.

\begin{figure}[h]
\centering
\includegraphics[width=\figsize\textwidth]{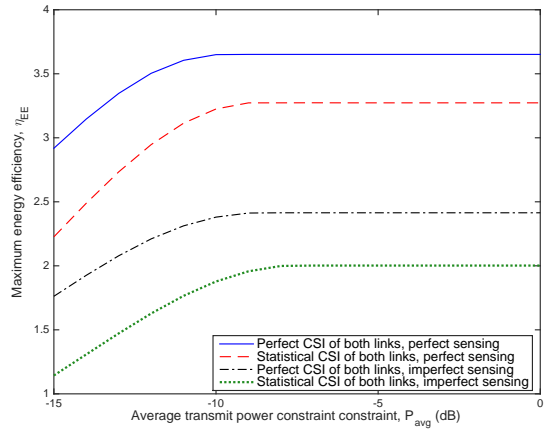}
\caption{Maximum EE $\eta_{\rm{EE}}$ vs. average transmit power constraint.}
\label{fig:EE_Pavg}
\end{figure}

In Fig. \ref{fig:EE_Pavg}, we plot maximum EE attained with the proposed optimal power allocation schemes as a function of the average transmit power constraint under perfect sensing with $\mathscr{P}_{\nid}=1$ and $\mathscr{P}_{\f}=0$) and imperfect sensing with $\mathscr{P}_{\nid}=0.8$ and $\mathscr{P}_{\f}=0.1$. We assume the availability of either perfect instantaneous CSI or statistical CSI of both transmission and interference links at the secondary transmitter. $Q_{\avg}$ is set to $-25$ dB. It is observed from the figure that when the optimal power allocation with perfect instantaneous CSI is applied, higher EE is achieved compared to the optimal power allocation with statistical CSI. More specifically, the power allocation scheme assuming perfect instantaneous CSI can exploit favorable channel conditions and higher transmission power is allocated to better channel, and hence a secondary user's power budget is more efficiently utilized compared to the power allocation scheme assuming statistical CSI in which the power levels do not change according to channel conditions. It is also seen that imperfect sensing decisions significantly affect the performance of secondary users, resulting in lower EE under both optimal power allocation strategies.

\begin{figure*}[htb]
\centering
\begin{subfigure}[b]{0.32\textwidth}
\centering
\includegraphics[width=\textwidth]{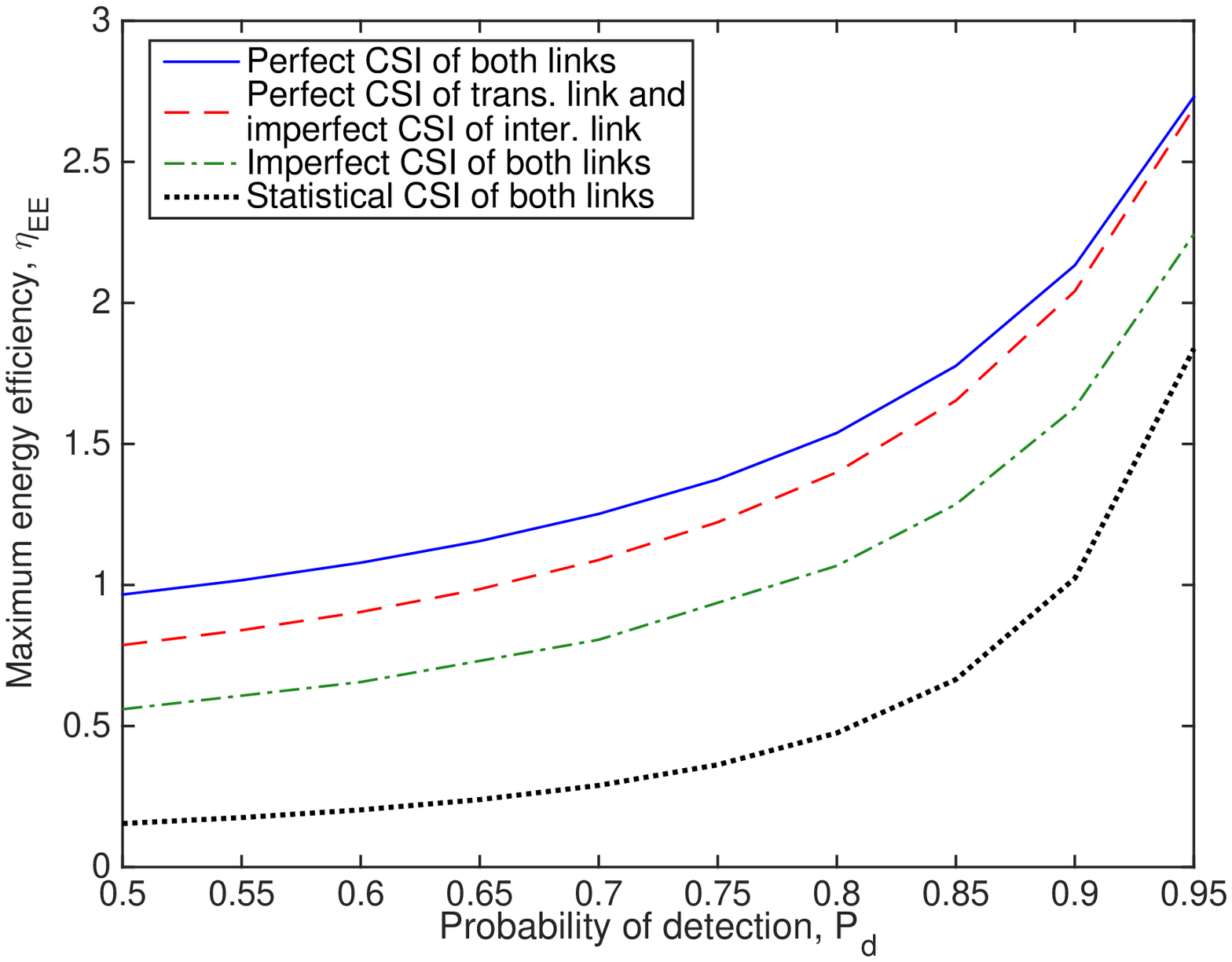}
\caption{$\eta_{\rm{EE}}$ vs. $\mathscr{P}_{\nid}$}
\end{subfigure}
\begin{subfigure}[b]{0.32\textwidth}
\centering
\includegraphics[width=\textwidth]{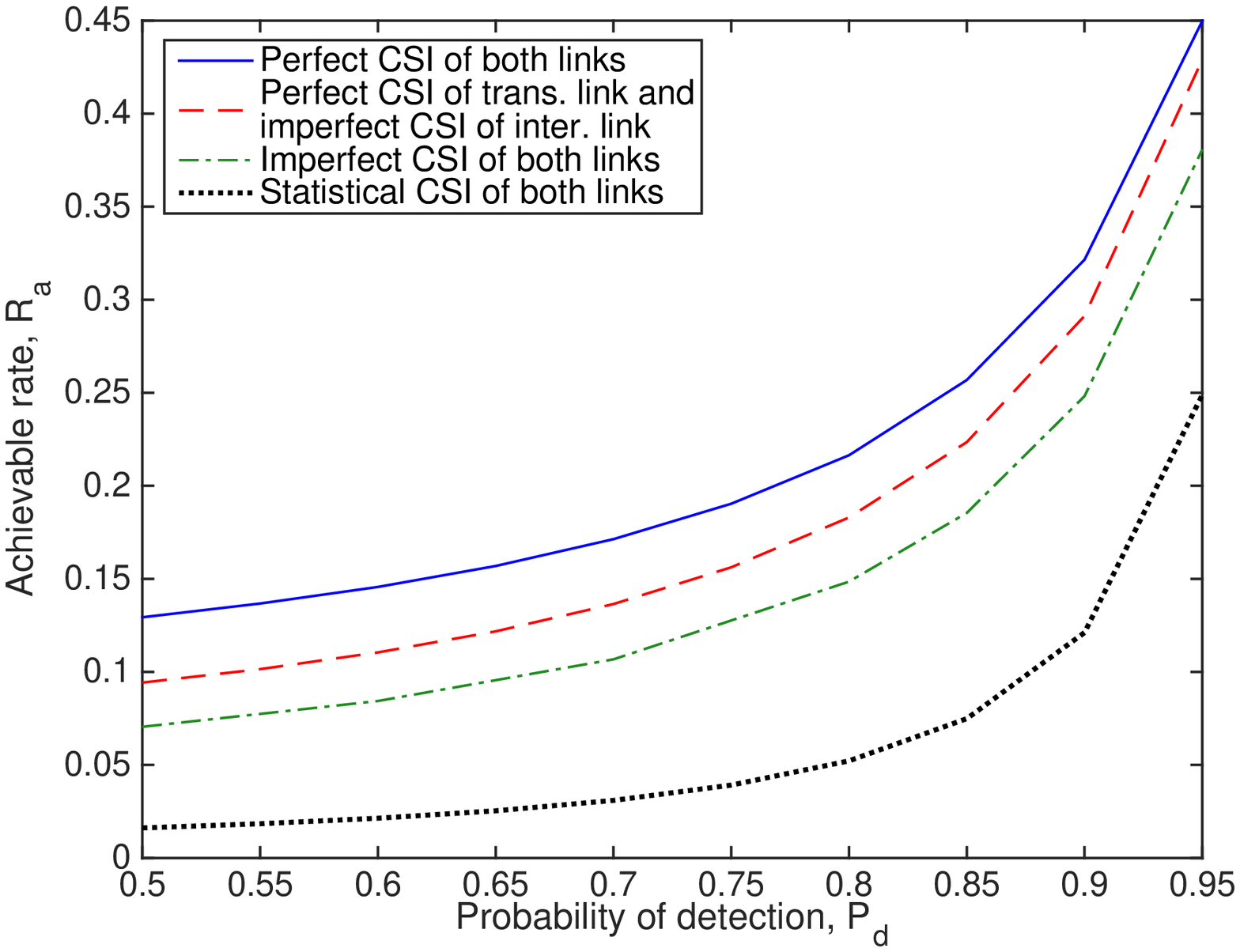}
\caption{$R_a$ vs. $\mathscr{P}_{\nid}$}
\end{subfigure}
\begin{subfigure}[b]{0.32\textwidth}
\centering
\includegraphics[width=\textwidth]{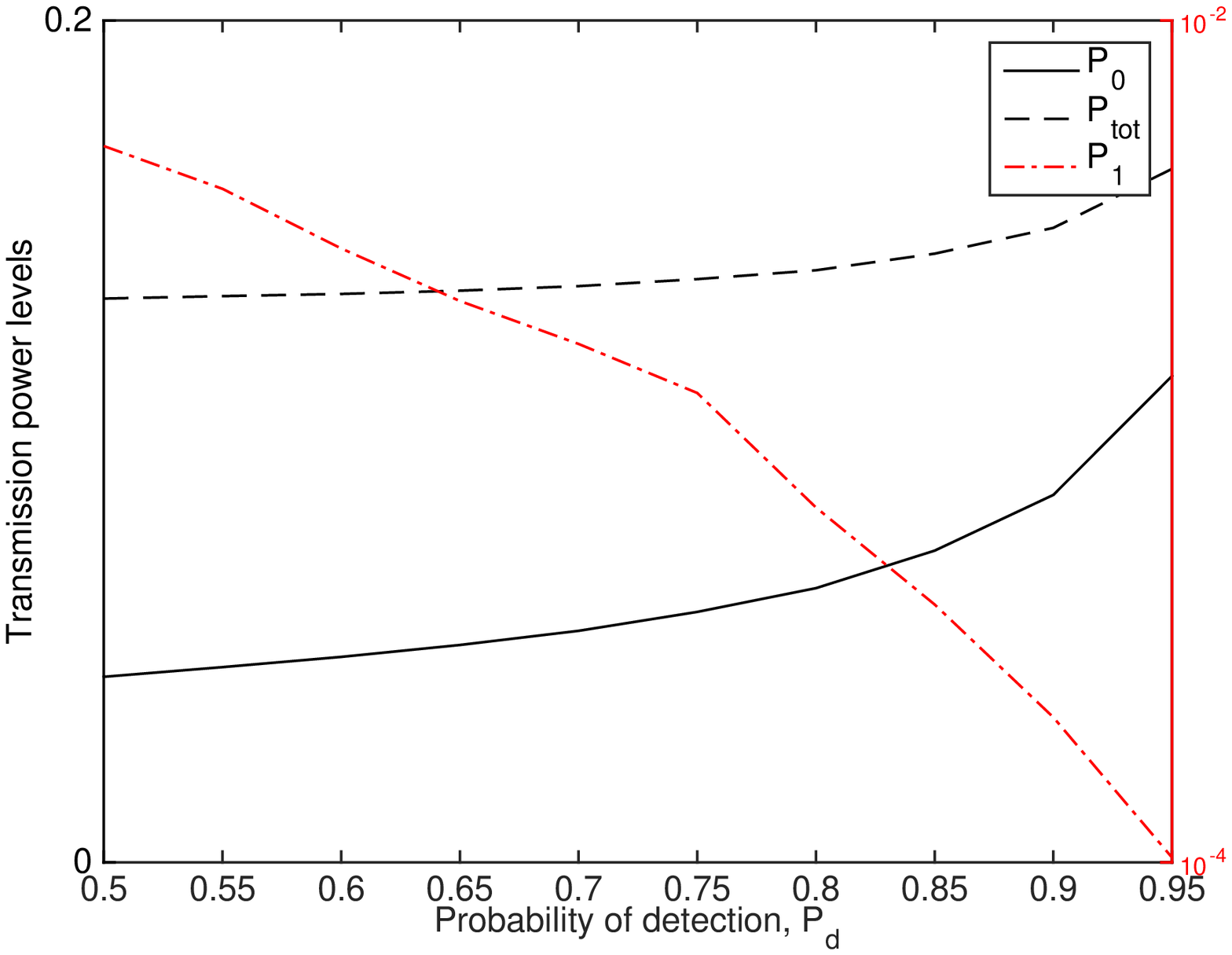}
\caption{$P_{\text{tot}}$, $P_0$ and $P_1$ vs. $\mathscr{P}_{\nid}$}
\end{subfigure}
\caption{\small{(a) Maximum achievable EE, $\eta_{\rm{EE}}$ vs. probability of detection, $\mathscr{P}_{\nid}$; (b) achievable rate maximizing EE, $R_a$ vs. $\mathscr{P}_{\nid}$; (c) optimal total transmission power, $P_{\text{tot}}$ and $P_0$, $P_1$ vs. $\mathscr{P}_{\nid}$.}}\label{fig:EE_Pd}
\end{figure*}

\begin{figure*}[htb]
\centering
\begin{subfigure}[b]{0.32\textwidth}
\centering
\includegraphics[width=\textwidth]{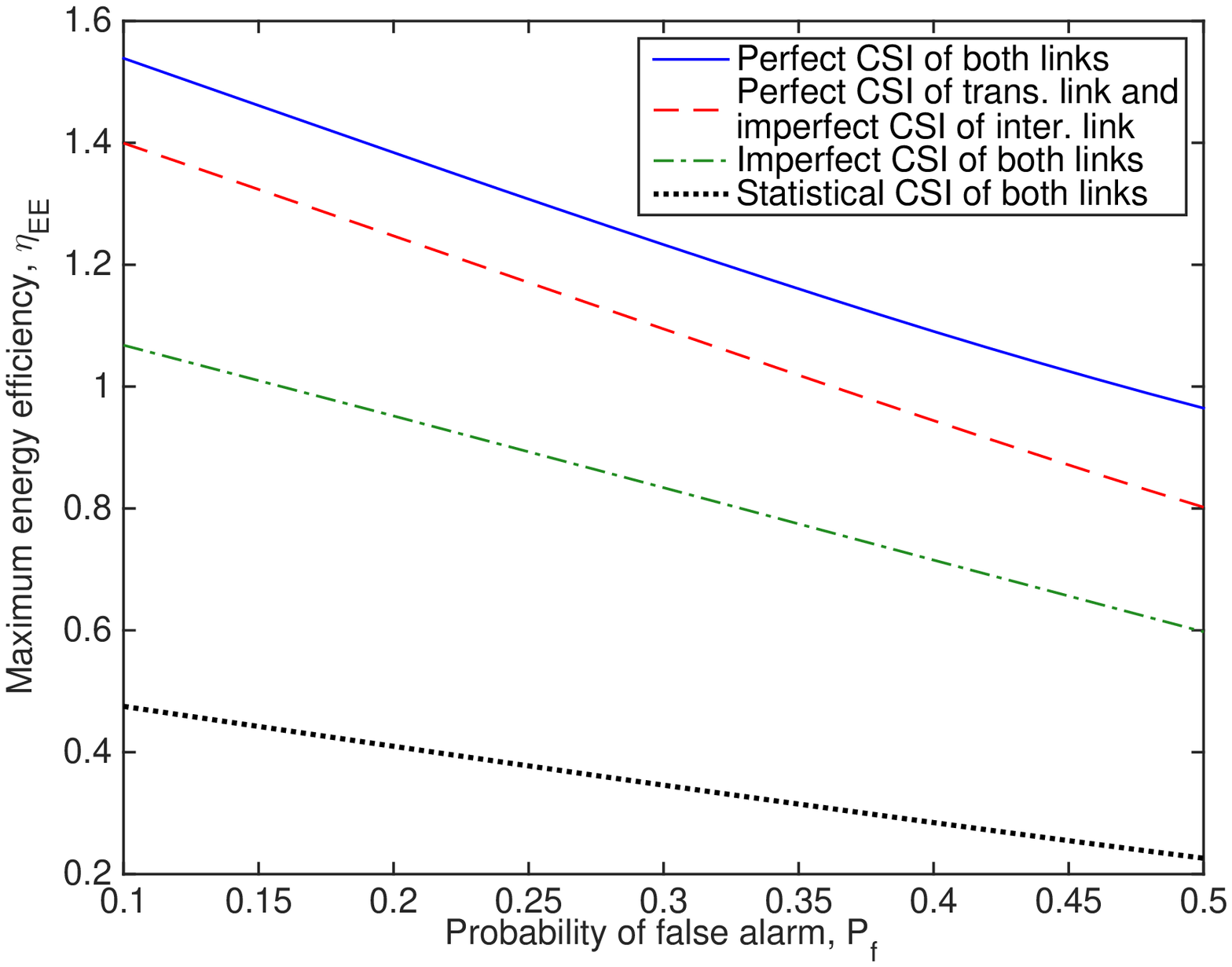}
\caption{$\eta_{\rm{EE}}$ vs. $\mathscr{P}_{\f}$}
\end{subfigure}
\begin{subfigure}[b]{0.32\textwidth}
\centering
\includegraphics[width=\textwidth]{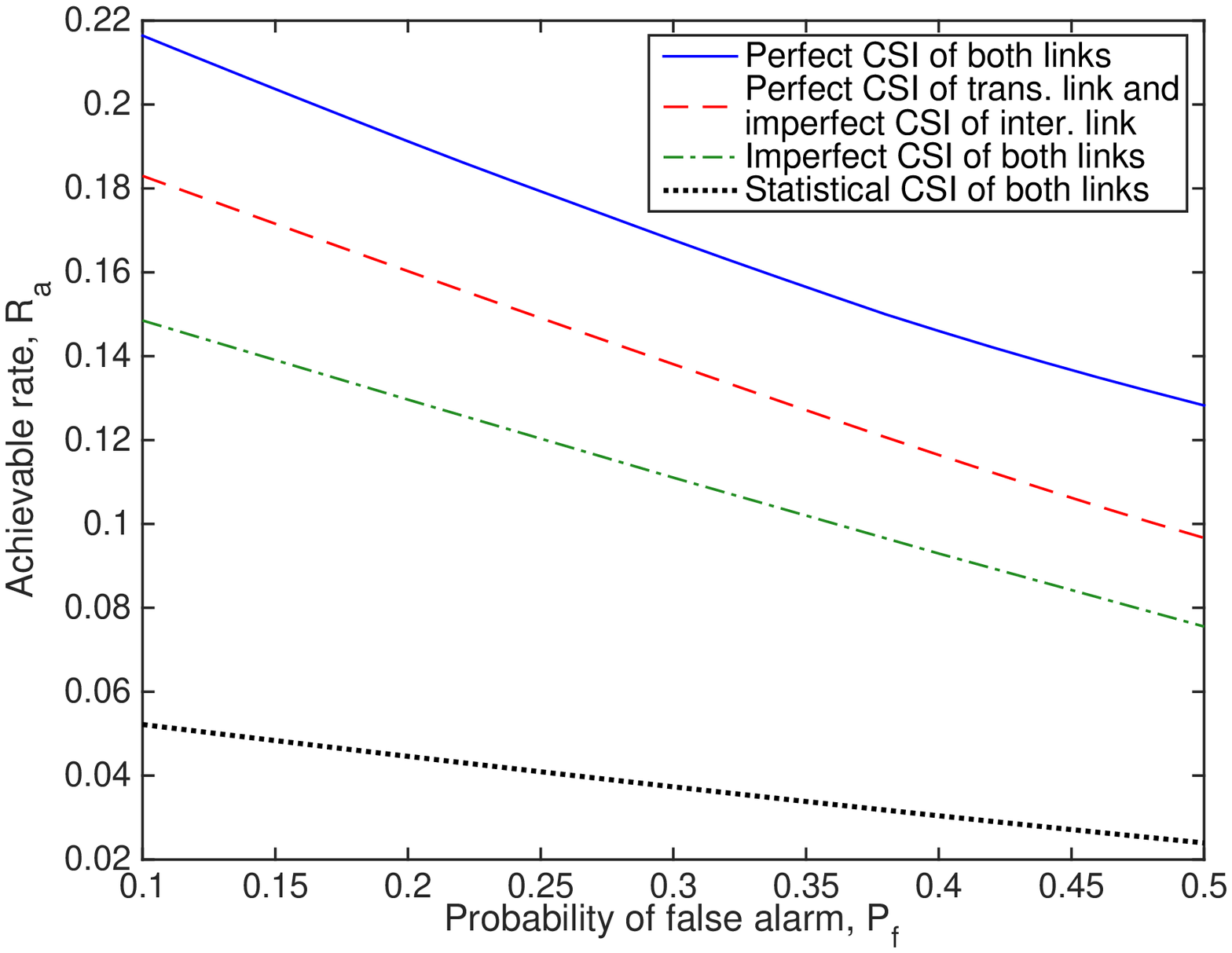}
\caption{$R_a$ vs. $\mathscr{P}_{\f}$}
\end{subfigure}
\begin{subfigure}[b]{0.32\textwidth}
\centering
\includegraphics[width=\textwidth]{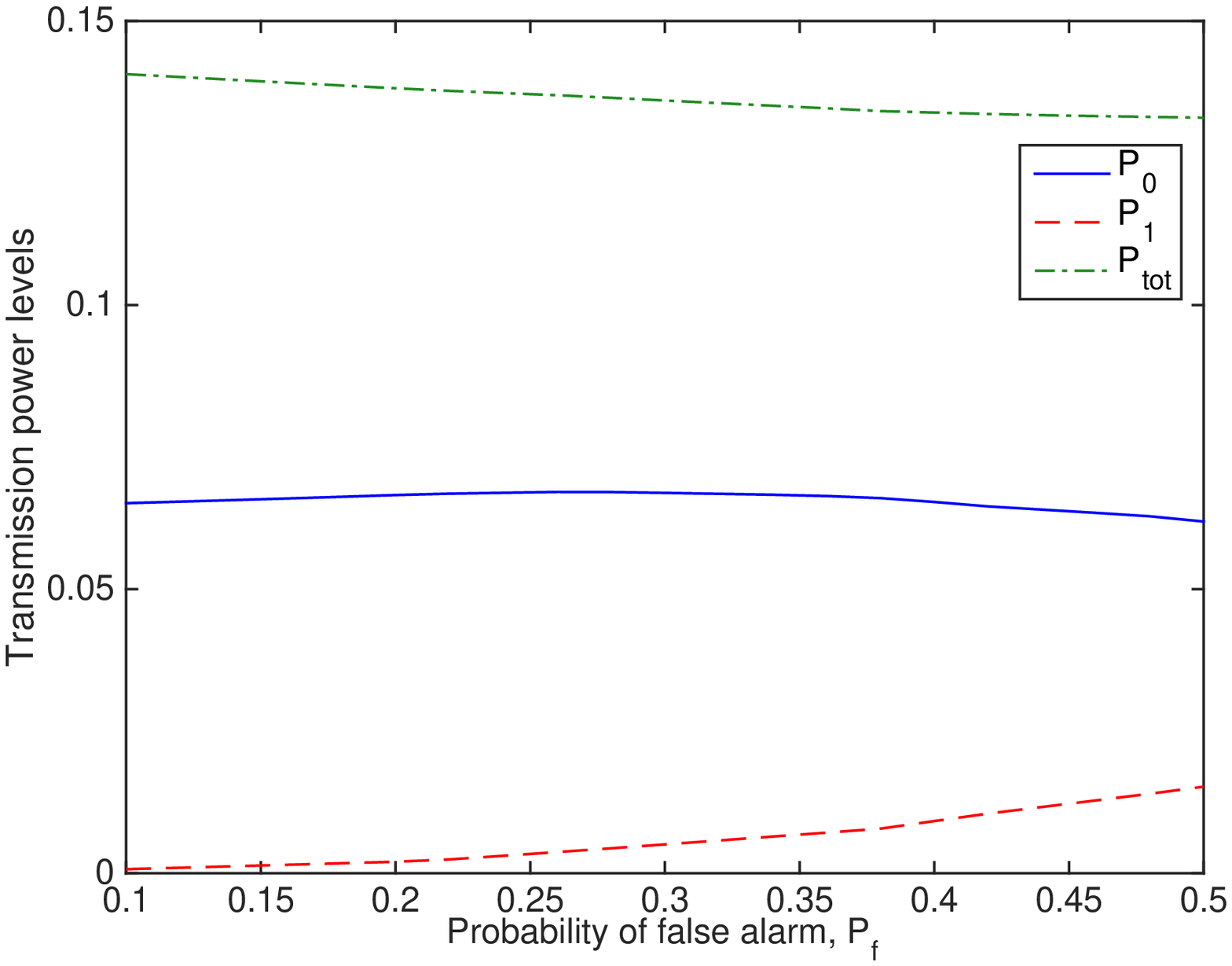}
\caption{$P_{\text{tot}}$, $P_0$ and $P_1$ vs. $\mathscr{P}_{\f}$}
\end{subfigure}
\caption{\small{(a) Maximum EE, $\eta_{\rm{EE}}$ vs. probability of false alarm, $\mathscr{P}_{\f}$; (b) achievable rate maximizing EE, $R_a$ vs. $\mathscr{P}_{\f}$; (c) optimal total transmission power, $P_{\text{tot}}$ and $P_0$, $P_1$ vs. $\mathscr{P}_{\f}$.}}\label{fig:EE_Pf}
\end{figure*}

In Fig. \ref{fig:EE_Pd}, we display maximum EE, achievable rate $R_a$, and optimal powers, $P_{\text{tot}}$, $P_0$ and $P_1$ as a function of detection probability, $\mathscr{P}_{\nid}$. It is assumed that peak transmit power constraints are $P_{\pk,0}=P_{\pk,1}=-4$ dB and average interference power constraint is $Q_{\avg}=-25$ dB. In addition, probability of false alarm, $\mathscr{P}_{\f}$ is set to $0.1$. We consider the power allocation schemes for the following four cases: (1) perfect CSI of both transmission and interference links; (2) perfect CSI of the transmission link and imperfect CSI of the interference link; (3) imperfect CSI of both transmission and interference links; and (4) only statistical CSI of both transmission and interference links. We only plot optimal powers, $P_{\text{tot}}$, $P_0$ and $P_1$ for the optimal power allocation with perfect instantaneous CSI of both links since the same trend is observed under the assumption of other CSI levels. As $\mathscr{P}_{\nid}$ increases, secondary users have more reliable sensing performance. Hence, secondary users experience miss detection events less frequently, which results in increased achievable rate. The transmission power under idle sensing decision, $P_0$ increases with increasing $\mathscr{P}_{\nid}$ while transmission power under busy sensing decision, $P_1$ decreases with increasing $\mathscr{P}_{\nid}$. Since achievable rate increases and total transmission power slightly increases, maximum EE of secondary users increases as sensing performance improves. It is also seen that the power allocation scheme with perfect instantaneous CSI of both links outperforms the other proposed power allocation strategies. Moreover, the performance of secondary users in terms of throughput and EE degrades gradually as we have less and less information regarding the transmission and interference links at the secondary transmitter.

Fig. \ref{fig:EE_Pf} shows maximum EE, achievable rate, $R_a$ and optimal powers, $P_{\text{tot}}$, $P_0$ and $P_1$ as a function of false alarm probability, $\mathscr{P}_{\f}$. We consider the same setting as in the previous figure. It is again assumed that $P_{pk,0}=P_{\pk,1} = -4$ dB and $Q_{\avg} = -8$ dB.  Since optimal powers maximizing EE show similiar trends as a function of $\mathscr{P}_{\f}$, we only plot the optimal power levels under the assumption of perfect instantaneous CSI of both transmission and interference links in Fig. \ref{fig:EE_Pf} (c). Probability of detection, $\mathscr{P}_{\nid}$ is chosen as $0.8$. As $\mathscr{P}_{\f}$ increases, channel sensing performance deteriorates. In this case, secondary users detect the channel as busy more frequently even if the channel is idle. Total transmission power maximizing EE slightly decreases with increasing $\mathscr{P}_{\f}$. In addition, since the available channel is not utilized efficiently,
 secondary users have smaller achievable rate, which leads to lower achievable EE. Again, the power allocation scheme with perfect instantaneous CSI of both links gives the best performance in terms of throughput and EE.

\begin{figure}[h]
\centering
\includegraphics[width=\figsize\textwidth]{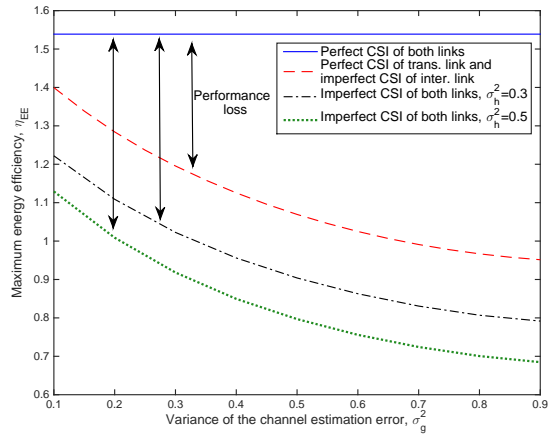}
\caption{Maximum EE $\eta_{\rm{EE}}$ vs. channel estimation error variance of the interference link, $\sigma)g^2$.}
\label{fig:EE_error_variance}
\end{figure}

In Fig. \ref{fig:EE_error_variance}, we display maximum EE as a function of channel estimation error variance of the interference link. Power allocation is employed by perfect CSI of transmission link and imperfect CSI of interference link, and imperfect CSI of both links with $\sigma_h^2=0.3$ and $\sigma_h^2=0.5$. We assume that $P_{pk,0}=P_{\pk,1} = -4$ dB and $Q_{\avg} = -25$ dB, and sensing is imperfect with $\mathscr{P}_{\nid}=0.8$ and $\mathscr{P}_{\f}=0.1$. The EE attained with the optimal power allocation assuming perfect CSI of both links is displayed as a baseline to compare the performance loss due to imperfect CSI of either interference link or of both transmission and interference links at the secondary transmitter. It is observed that EE of secondary users decreases as the variance of the channel estimation error in the interference link, $\sigma_g^2$, increases and hence the channel estimate becomes less accurate. The secondary users even have lower EE when CSI of both links are imperfectly known. Therefore, accurate estimation of both transmission and interference links is crucial in order to achieve better EE.

\begin{figure}[h]
\centering
\includegraphics[width=\figsize\textwidth]{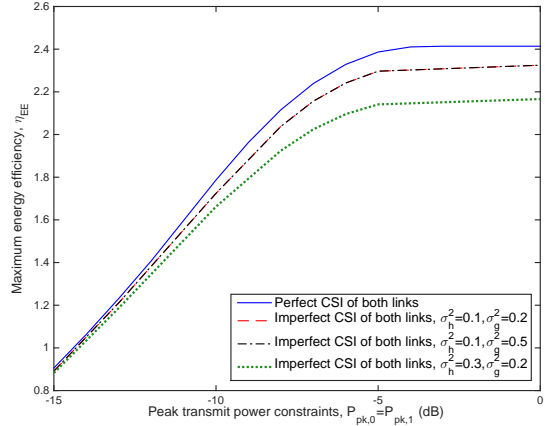}
\caption{Maximum EE $\eta_{\rm{EE}}$ vs. peak transmit power constraints, $P_{\pk,0}=P_{\pk,1}$.}
\label{fig:EE_Ppk}
\end{figure}

Fig. \ref{fig:EE_Ppk} shows the maximum EE as a function of the peak transmit power constraints, $P_{\pk,0}=P_{\pk,1}$ under imperfect sensing result (i.e., when $\mathscr{P}_{\nid}=0.8$ and $\mathscr{P}_{\f}=0.1$). We consider the optimal power allocation schemes assuming either perfect CSI of both links or imperfect CSI of both links with $\sigma_h^2=0.1$ and $\sigma_g^2=0.2$, $\sigma_h^2=0.1$ and $\sigma_g^2=0.5$, and $\sigma_h^2=0.3$ and $\sigma_g^2=0.2$. Average interference constraint, $Q_{\avg}$ is set to $-10$ dB.  When channel estimation error of the transmission link increases from $0.1$ and $0.3$ keeping $\sigma_g^2=0.2$, EE of secondary users decreases more compared to the case when the channel estimation error of the interference link increases from $0.2$ to $0.5$ while $\sigma_h^2=0.1$ since the average interference constraint is loose, and imperfect CSI of the interference link only slightly affects the performance. Also, as $P_{\pk}$ increases, EE of secondary users first increases and then stays constant since average transmission power reaches to the value that maximizes EE. Therefore, further increasing $P_{\pk}$ does not provide any EE improvement.

\begin{figure}[h]
\centering
\includegraphics[width=\figsize\textwidth]{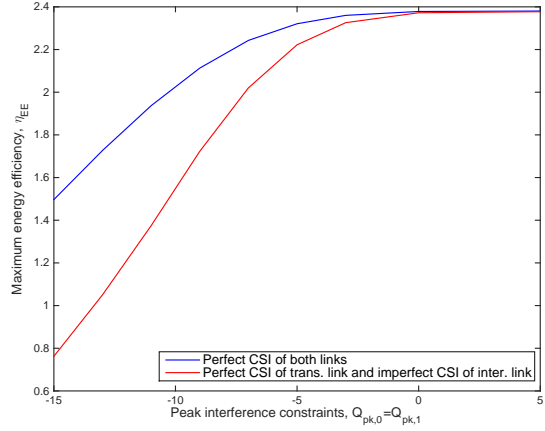}
\caption{Maximum EE $\eta_{\rm{EE}}$ vs. peak interference power constraints, $Q_{\pk,0}=Q_{\pk,1}$.}
\label{fig:EE_Qpk}
\end{figure}

In Fig. \ref{fig:EE_Qpk}, we plot the maximum EE as a function of the peak interference power constraints, $Q_{\pk,0}=Q_{\pk,1}$. It is assumed that $\mathscr{P}_{\nid}=0.8$ and $\mathscr{P}_{\f}=0.1$. We consider that either perfect instantaneous CSI of both links or perfect CSI of transmission link and imperfect CSI of interference link is available at the secondary transmitter. We set $P_{\avg}= -10$ dB, outage thresholds $\xi_0=\xi_1 = 0.1$ and $\sigma_g^2=0.1$. It is seen that EE curves under both cases first increase with increasing interference power constraints, and then level and approach the same value due to average transmit power constraint. Also, the availability of only imperfect CSI of the interference link deteriorates the system performance and leads to lower EE compared to that with perfect CSI of the interference link.

\section{Conclusion}\label{sec:conc}
In this paper, we determine energy-efficient power allocation schemes for cognitive radio systems subject to peak/average transmit power constraints and peak/average interference power constraint in the presence of sensing errors and different levels of CSI regarding the transmission and interference links. A low-complexity algorithm based on Dinkelbach's method is proposed to iteratively solve the power allocation that maximizes EE. It is shown that power allocation schemes depend on sensing performance through detection and false alarm probabilities, transmission link between secondary transmitter and secondary receiver, and interference link between the secondary transmitter and the primary receiver. Numerical results reveal interesting relations and tradeoffs. For instance, it is shown that maximum achievable EE increases with increasing $\mathscr{P}_{\nid}$ and decreases with increasing $\mathscr{P}_{\f}$. Imperfect CSI of the transmission and interference links significantly degrades the performance of secondary users in terms of EE. Therefore, accurate estimation of the transmission and interference links is of great importance in order to obtain higher EE. Moreover, under the same average interference constraint, secondary users' transmission subject to peak transmit constraint achieves smaller achievable EE than that under average transmit power constraint.

\appendix
\subsection{Proof of Proposition \ref{prop:1}} \label{appendix1}
We first write the achievable rate of secondary users in terms of mutual information between the received and transmitted signals given the sensing decision as follows:
\begin{align} \label{eq:mutual_info_sensing}
R_{G}=\frac{T-\tau}{T}I(x;y|h,\hH)=\frac{T-\tau}{T}\bigg[&\Pr\{\hH_0\}I(x_0;y|h,\hH_0)+\Pr\{\hH_1\}I(x_1;y|h,\hH_1)\bigg],
\end{align}
where $I(x_k;y|h,\hH_k)$ for $k \in \{0,1\}$ can be further expressed as
\begin{align} \label{eq:mutual_info_imperfect}
I(x_k;y|h,\hH_k)= \E_{x_k,y,g,h}\bigg\{\log\bigg(\frac{f(y|x_k,h,\hH_k)}{f(y|h,\hH_k)}\bigg)\bigg\}.
\end{align}
The above conditional distribution, $f(y|x_k,h,\hH_k)$, is determined through the input-output relation in (\ref{eq:received_signa_BSl}) as follows:
\begin{align} 
f(y|x_k,h,\hH_k)=&\frac{\Pr\{\mH_0|\hH_k\}}{\pi N_0}\rme^{-\frac{|y-hx_k|^2}{N_0}} + \frac{\Pr\{\mH_1| \hH_k\}}{\pi (N_0+\sigma_s^2)}\rme^{-\frac{|y-hx_k|^2}{N_0+\sigma_s^2}}
\end{align}
with variance
\begin{align}
\E\{|y|^2|x_k,h,\hH_k\}=N_0 + \Pr\{\mH_1|\hH_k\}\sigma_s^2.
\end{align}
Also, assuming Gaussian distributed input, the conditional distribution, $f(y|h,\hH_k)$ in (\ref{eq:mutual_info_imperfect}) is given by
\begin{align} 
f(y|h,\hH_k)=&\frac{\Pr\{\mH_0|\hH_k\}}{\pi (N_0+P_k(g,h)|h|^2)}\rme^{-\frac{|y|^2}{N_0+P_k(g,h)|h|^2}} + &\frac{\Pr\{\mH_1|\hH_k\}}{\pi (N_0+P_k(g,h)|h|^2+\sigma_s^2)}\rme^{-\frac{|y|^2}{N_0+P_k(g,h)|h|^2+\sigma_s^2}}
\end{align}
with variance
\begin{align}
\E\{|y|^2|h,\hH_k\}=N_0 + P_k(g,h)|h|^2 + \Pr\{\mH_1|\hH_k\}\sigma_s^2.
\end{align}
Above, it is seen that the conditional distributions of the received signal $y$ given sensing decisions become a mixture of Gaussian distributions due to channel sensing errors. Therefore, there is no closed form expression for mutual information in (\ref{eq:mutual_info_imperfect}). However, we can still find a closed-form lower bound for the achievable rate expression by following the steps in \cite[pp. 938-939]{medard} and replacing the additive disturbance $w$ in (\ref{eq:disturbance}) with the worst-case Gaussian noise with the same variance as follows:
\begin{align}
I(x_k;y|h,\hH_k) \geq \E_{g,h}\bigg\{\log\bigg(1+\frac{P_k(g,h)|h|^2}{E\{|w|^2|\hH_k\}}\bigg)\bigg\}
\end{align}
where
\begin{align}
E\{|w|^2|\hH_k\}=N_0 + \Pr\{\mH_1|\hH_k\}\sigma_s^2.
\end{align}
Inserting these lower bounds into (\ref{eq:mutual_info_sensing}), we have obtained the achievable rate expression in (\ref{eq:lower_bound_R}).  \hfill $\square$

\subsection{Proof of Theorem \ref{teo:upper_bound}} \label{appendix2}
We first express $I_0(x_0;y|h,\hH_0)$ in (\ref{eq:I_0_exact}) given at the top of page, where $c_1=N_0 + \sigma_s^2$ and $c_2= N_0$.
\begin{figure*}
\begin{equation}
\small
\begin{split} \label{eq:I_0_exact}
&I_0(x_0;y|h,\hH_0)=\Big(\frac{T-\tau}{T}\Big)\Bigg[\int_{h}\bigg(\int_{-\infty}^{\infty}\sum_{i=1}^{2}\frac{\Pr(\mH_i|\hH_0)}{\pi c_i} \rme^{-\frac{|y-x_0 h|^2}{c_i}}\log \Big(\sum_{i=1}^{2}\frac{\Pr(\mH_i|\hH_0)}{\pi c_i} \rme^{-\frac{|y-x_0 h|^2}{c_i}}\Big)dy \bigg)f_{h}(h)dh \\& - \int_{g}\Big(\int_{h}\bigg(\int_{-\infty}^{\infty}\sum_{i=1}^2 \frac{\Pr(\mH_i|\hH_0)}{\pi (c_i + |h|^2P_0(g,h))}\rme^{-\frac{|y|^2}{c_i+|h|^2P_0(g,h)}}\log \Big(\sum_{i=1}^2 \frac{\Pr(\mH_i|\hH_0)}{\pi (c_i + |h|^2P_0(g,h))}\rme^{-\frac{|y|^2}{c_i+|h|^2P_0(g,h)}}\Big)dy\bigg) f_{h}(h)dh \Big) f_{g}(g)dg \Bigg]
\end{split}
\normalsize
\end{equation}
\hrule
\end{figure*}
\begin{figure*}
\begin{equation}
\small
\begin{split} \label{eq:I_0_upper}
&I_0(x_0;y|h,\hH_0)\le \Big(\frac{T-\tau}{T}\Big)\Bigg[\int_{h}\bigg(\int_{-\infty}^{\infty}\sum_{i=1}^{2}\frac{\Pr(\mH_i|\hH_0)}{\pi c_i} \rme^{-\frac{|y-x_0 h|^2}{c_i}}\log \Big(\sum_{i=1}^{2}\frac{\Pr(\mH_i|\hH_0)}{\pi c_i} \rme^{-\frac{|y-x_0 h|^2}{c_1}}\Big)dy \bigg)f_{h}(h)dh \\& - \int_{g}\Big(\int_{h}\bigg(\int_{-\infty}^{\infty}\sum_{i=1}^2 \frac{\Pr(\mH_i|\hH_0)}{\pi (c_i + |h|^2P_0(g,h))}\rme^{-\frac{|y|^2}{c_i+|h|^2P_0(g,h)}}\log \Big(\sum_{i=1}^2 \frac{\Pr(\mH_i|\hH_0)}{\pi (c_i + |h|^2P_0(g,h))}\rme^{-\frac{|y|^2}{c_2+|h|^2P_0(g,h)}}\Big)dy\bigg) f_{h}(h)dh \Big) f_{g}(g)dg \Bigg]
\end{split}
\normalsize
\end{equation}
\hrule
\end{figure*}
Then, we obtain the upper bound in (\ref{eq:I_0_upper}) by using the following inequalities:
\begin{align}
\rme^{-\frac{|y|^2}{c_1 + |h|^2 P_0(g,h)}} &\geq \rme^{-\frac{|y|^2}{c_i + |h|^2 P_0(g,h)}} \geq  \rme^{-\frac{|y|^2}{c_2 + |h|^2 P_0(g,h)}} \\
\rme^{-\frac{|y-h x_0|^2}{c_1}} &\geq \rme^{-\frac{|y- h x_0|^2}{c_i}} \geq  \rme^{-\frac{|y - h x_0|^2}{c_2}}.
\end{align}
Evaluating the integrals in (\ref{eq:I_0_upper}) yields the following upper bound:
\begin{equation}
\small
\begin{split} \label{eq:I_0_inequal}
&I_0(x_0;y|h,\hH_0) \le \Big(\frac{T-\tau}{T}\Big)\Bigg[ \E_{g,h}\bigg\{\sum_{k=1}^2\log \Bigg(\frac{\sum_{i=1}^2\frac{\Pr (\mH_i|\hH_0)}{c_i}}{\sum_{i=1}^2\frac{\Pr (\mH_i | \hH_0)}{c_i + |h|^2P_0(g,h)}}\Bigg)\bigg\} -\bigg(\frac{N_0 + \Pr (H_1 | \hH_0)\sigma_s^2}{N_0 + \sigma_s^2}\bigg) + \E_{g,h}\bigg\{1+\frac{\Pr (H_1 | \hH_0)\sigma_s^2}{N_0 + |h|^2P_0(g,h)}\bigg\} \Bigg]
\end{split}
\normalsize
\end{equation}
Following the similar steps, we obtain the upper bound for $I_1(x_1;y|h,\hH_1)$ as follows:
\begin{equation}
\small
\begin{split}  \label{eq:I_1_inequal}
&I_1(x_1;y|h,\hH_1) \le \Big(\frac{T-\tau}{T}\Big) \Bigg[ \E_{g,h}\bigg\{\sum_{k=1}^2\log \Bigg(\frac{\sum_{i=1}^2\frac{\Pr (\mH_i|\hH_1)}{c_i}}{\sum_{i=1}^2\frac{\Pr (\mH_i | \hH_1)}{c_i + |h|^2P_1(g,h)}}\Bigg)\bigg\} -\bigg(\frac{N_0 + \Pr (H_1 | \hH_1)\sigma_s^2}{N_0 + \sigma_s^2}\bigg) + \E_{g,h}\bigg\{1+\frac{\Pr (H_1 | \hH_1)\sigma_s^2}{N_0 + |h|^2P_1(g,h)}\bigg\}  \Bigg]
\end{split}
\normalsize
\end{equation}
Inserting the inequalities in (\ref{eq:I_0_inequal}) and (\ref{eq:I_1_inequal}) into (\ref{eq:mutual_info_sensing}) and subtracting $R_a$ in (\ref{eq:lower_bound_R}) give the desired result in (\ref{eq:upper_bound1}). \hfill $\square$

\subsection{Proof of Theorem \ref{teo:1}} \label{appendix3}

The optimization problem is quasiconcave since achievable rate $R_a$ is concave in transmission powers and total power consumption $P_{\text{tot}}(P_0,P_1)$ is both affine and positive, and hence the level sets $S_{\alpha}=\{(P_0, P_1) : \eta_{\rm{EE}}\geq \alpha\}=\{\alpha P_{\text{tot}}(P_0,P_1)- R_a \le 0\}$ are convex for any $\alpha \in \mathbb{R}$. Since quasiconcave functions have more than one local maximum, local maximum does not always guarantee the global maximum. Therefore, standard convex optimization algorithms cannot be directly used. Hence, iterative power allocation algorithm based on Dinkelbach's method \cite{dinkelbach} is employed to solve the quasiconcave EE maximization problem by formulating the equivalent parameterized concave problem as follows:
\begin{align}\label{eq:parametrized_EE}
&\max_{
\substack{P_0(g,h) \\ P_1(g,h)}}\Big\{\E_{g,h}\big\{R\big(P_0(g,h),P_1(g,h)\big)\big\}-\alpha (\E_{g,h}\{\Pr\{\hH_0\}P_0(g,h)+\Pr\{\hH_1\}P_1(g,h)\}\!+\!P_{\mc})\Big\} \\
\label{eq:Pavg_cons}
&\text{subject to} \hspace{0.4cm}\E_{g,h}\{\Pr\{\hH_0\}\,P_0(g,h)  + \Pr\{\hH_1\}\,P_1(g,h)\}  \le P_{\avg} \\ \label{eq:Qavg_cons}
&\hspace{2.1cm}\E_{g,h}\{\big[(1-\mathscr{P}_{\nid})\,P_0(g,h) + \mathscr{P}_{\nid} \,P_1(g,h)\big]|g|^2\} \le Q_{\avg}\\
&\hspace{2.1cm}P_0(g,h)\geq 0 , P_1(g,h)\geq 0,
\end{align}
\normalsize
where $\alpha$ is a nonnegative parameter. At the optimal value of $\alpha^*$, solving the EE maximization problem in (\ref{eq:EE_avg1}) is equivalent to solving the above parametrized concave problem if and only if the following condition is satisfied
\begin{equation}
\begin{split} \label{eq:cond_opt}
F(\alpha^*)=\E_{g,h}\big\{R\big(P_0(g,h),P_1(g,h)\big)\big\} -\alpha^* \Big(\E_{g,h}\{\Pr\{\hH_0\}P_0(g,h)+\!\Pr\{\hH_1\}P_1(g,h)\}+P_{\mc}\Big)=0.
\end{split}
\end{equation}
The detailed proof of the above condition is available in \cite{dinkelbach}. Since the problem in (\ref{eq:parametrized_EE}) is concave, the optimal power values are obtained by forming the Lagrangian as follows:
\begin{equation}
\small
\begin{split}
&L(P_0,P_1,\lambda_1,\nu_1,\alpha)=\E_{g,h}\big\{R\big(P_0(g,h),P_1(g,h)\big)\big\}-\alpha (\E_{g,h}\{\Pr\{\hH_0\}P_0(g,h)+\Pr\{\hH_1\}P_1(g,h)\}\!+\!P_{\mc})\\&-\lambda_1 (\E_{g,h}\{\Pr\{\hH_0\}\,P_0(g,h)  + \Pr\{\hH_1\}\,P_1(g,h)\} -P_{\avg} )-\nu_1 (\E_{g,h}\{\big[(1-\mathscr{P}_{\nid})\,P_0(g,h) + \mathscr{P}_{\nid} \,P_1(g,h)\big]|g|^2\} -Q_{\avg}),
\end{split}
\end{equation}
\normalsize
where $\lambda_1$ and $\nu_1$ are nonnegative Lagrange multipliers. According to the Karush-Kuhn-Tucker (KKT) conditions, the optimal values of $P_0^*(g,h)$ and $P_1^*(g,h)$ satisfy the following equations:
\begin{align} \label{eq:P0_Lagrange1}
&\frac{\frac{T-\tau}{T}\Pr\{\hH_0\}|h|^2\log_2e}{N_0+\Pr(\mH_1|\hH_0)\sigma_s^2+P_0^*(g,h)|h|^2}-(\lambda_1+\alpha) \Pr\{\hH_0\} -\nu_1 |g|^2(1-\mathscr{P}_{\nid})=0 \\
&\frac{\frac{T-\tau}{T}\Pr\{\hH_1\}|h|^2\log_2e}{N_0\!+\!\Pr(\mH_1\!|\hH_1\!)\sigma_s^2+P_1^*(g,\!h)|h|^2}-(\lambda_1\!+\!\alpha)  \Pr\{\hH_1\!\} -\nu_1 |g|^2\mathscr{P}_{\nid}\!=\!0  \label{eq:P1_Lagrange1} \\
&\lambda_1 (\E\{\Pr\{\hH_0\}\,P_0^*(g,h)  + \Pr\{\hH_1\}\,P_1^*(g,h)\}  - P_{\avg}) =0\\
& \nu_1 (\E\{[(1-\mathscr{P}_{\nid})\,P_0^*(g,h) + \mathscr{P}_{\nid} \,P_1^*(g,h)]|g|^2\} -Q_{\avg})=0 \\
& \lambda_1 \geq 0, \nu_1 \geq 0.
\end{align}
Solving equations (\ref{eq:P0_Lagrange1}) and (\ref{eq:P1_Lagrange1}), yield the optimal power values in (\ref{eq:opt_P0_avg1}) and (\ref{eq:opt_P1_avg1}), respectively. \hfill $\square$

\end{spacing}

\end{document}